\documentclass[aps,prd,reprint,superscriptaddress]{revtex4-2}

\usepackage{graphicx}
\usepackage{hyperref}
\usepackage{float}
\usepackage{makecell}
\usepackage{lineno}
\usepackage{amsmath}
\usepackage{array}
\usepackage{booktabs}

\begin{document}

\title{Study of $e^+e^-\to\omega X(3872)$ and $\gamma X(3872)$ from 4.66 to 4.95~GeV}

\author{
M.~Ablikim$^{1}$, M.~N.~Achasov$^{4,c}$, P.~Adlarson$^{75}$, O.~Afedulidis$^{3}$, X.~C.~Ai$^{80}$, R.~Aliberti$^{35}$, A.~Amoroso$^{74A,74C}$, Q.~An$^{71,58,a}$, Y.~Bai$^{57}$, O.~Bakina$^{36}$
% ...
\\
\vspace{0.2cm}
(BESIII Collaboration)\\
\vspace{0.2cm} {\it
$^{1}$ Institute of High Energy Physics, Beijing 100049, People's Republic of China\\
% ...
$^{3}$ Bochum  Ruhr-University, D-44780 Bochum, Germany\\
$^{4}$ Budker Institute of Nuclear Physics SB RAS (BINP), Novosibirsk 630090, Russia\\
% ...
$^{35}$ Johannes Gutenberg University of Mainz, Johann-Joachim-Becher-Weg 45, D-55099 Mainz, Germany\\
$^{36}$ Joint Institute for Nuclear Research, 141980 Dubna, Moscow region, Russia\\
% ...
$^{57}$ Southeast University, Nanjing 211100, People's Republic of China\\
$^{58}$ State Key Laboratory of Particle Detection and Electronics, Beijing 100049, Hefei 230026, People's Republic of China\\
% ...
$^{71}$ University of Science and Technology of China, Hefei 230026, People's Republic of China\\
% ...
$^{74}$ University of Turin and INFN, (A)University of Turin, I-10125, Turin, Italy; (B)University of Eastern Piedmont, I-15121, Alessandria, Italy; (C)INFN, I-10125, Turin, Italy\\
$^{75}$ Uppsala University, Box 516, SE-75120 Uppsala, Sweden\\
% ...
$^{80}$ Zhengzhou University, Zhengzhou 450001, People's Republic of China\\

\vspace{0.2cm}
$^{a}$ Deceased\\
% ...
$^{c}$ Also at the Novosibirsk State University, Novosibirsk, 630090, Russia\\
% ...
}
}

\begin{abstract}
Using data samples with an integrated luminosity of $4.5~\text{fb}^{-1}$ collected by the BESIII detector at center-of-mass energies ranging from 4.66 to 4.95~GeV, we study the processes of $e^+e^-\to\omega X(3872)$ and $e^+e^-\to\gamma X(3872)$. With the $e^+e^-\to\omega X(3872)$ process, the branching fraction ratio $R\equiv\frac{\mathcal{B}(X(3872)\to\gamma J/\psi)}{\mathcal{B}(X(3872)\to\pi^+\pi^- J/\psi)}$  is measured to be $0.38\pm0.20_\text{stat.}\pm0.01_\text{syst.}$ ($R< 0.83$ at 90\% confidence level). In addition, we measure the ratio of the average cross section of $e^+e^-\to\omega X(3872)$ to $e^+e^-\to\omega \chi_{c1}(\omega\chi_{c2})$ to be $\sigma_{\omega X(3872)}/\sigma_{\omega\chi_{c1}}~(\sigma_{\omega X(3872)}/\sigma_{\omega\chi_{c2}})=5.2\pm1.0_\text{stat.}\pm1.9_\text{syst.}~ (5.5\pm1.1_\text{stat.}\pm2.4_\text{syst.})$.  Finally, we search for the process of $e^+e^-\to\gamma X(3872)$, and no obvious signal is observed. The upper limit on the ratio of the average cross section of $e^+e^-\to\gamma X(3872)$ to $e^+e^-\to\omega X(3872)$ is set as $\sigma_{\gamma X(3872)}/\sigma_{\omega X(3872)}<0.23$ at 90\% confidence level.

\end{abstract}


\maketitle

\section{Introduction}
In 2003, the first charmonium-like state $X(3872)$ was  discovered by the Belle experiment in the $B^\pm\to K^\pm\pi^+\pi^-J/\psi$ process~\cite{Belle:2003nnu}. The mass of $X(3872)$ being right at the $D^0\bar{D}^{*0}$ threshold, along with its remarkably narrow width, suggests that the $X(3872)$ is not a conventional charmonium state.  
Its discovery has triggered extensive discussions, and interpretations as a $D^0\bar{D}^{*0}$ molecule~\cite{Swanson:2003tb,Guo:2017jvc}, a tetraquark state~\cite{Maiani:2004vq,Chen:2016qju}, or a mixture of a $D^0\bar{D}^{*0}$ molecule and a $\chi_{c1}(2P)$~\cite{Suzuki:2005ha,Matheus:2009vq,Ortega:2009hj} have been proposed. Up to now, there is, 
however, no consensus on the nature of the $X(3872)$.

After two decades, our knowledge on the $X(3872)$ has significantly advanced. Its quantum numbers $J^{PC}=1^{++}$~\cite{LHCb:2015jfc} and isospin $I=0$~\cite{BaBar:2004cah,Belle:2011vlx} were determined and several decay modes, including $\pi^+\pi^-J/\psi$~\cite{Belle:2011vlx}, $\omega J/\psi$~\cite{BaBar:2010wfc}, $D^0\bar{D}^{*0}$~\cite{BaBar:2007cmo}, $\gamma J/\psi$~\cite{Belle:2011wdj}, $\pi^0\chi_{c1}$~\cite{BESIII:2019esk}, were observed.  Among these, the radiative decay is particularly sensitive to its wave function~\cite{Cincioglu:2019gzd,Cardoso:2014xda,Guo:2014taa,Badalian:2015dha,Ke:2011jf,Badalian:2012jz,Nielsen:2010ij,Wang:2010ej,Mehen:2011ds,DeFazio:2008xq,Dubnicka:2011mm,Yu:2023nxk}. The BESIII experiment has measured the branching fraction ratio $R\equiv \frac{\mathcal{B}(X(3872)\to\gamma J/\psi)}{\mathcal{B}(X(3872)\to\pi^+\pi^- J/\psi)}=0.79\pm0.28$ through the $e^+e^-\to\gamma X(3872)$ process~\cite{BESIII:2020nbj}, which is consistent within $3\sigma$ with Belle's result of $R=0.21\pm0.06$, obtained from the decay $B^\pm\to K^\pm X(3872)$~\cite{Belle:2011vlx,Belle:2011wdj}. Compared to the $e^+e^-\to\gamma X(3872)$ process, the recently observed   $e^+e^-\to\omega X(3872)$ process by  BESIII~\cite{BESIII:2022bse} provides a lower background environment for measuring $R$.

In addition to the decays of $X(3872)$, investigating its production also provides valuable insight into its properties~\cite{Belle:2003nnu,D0:2004zmu,BaBar:2004cah,BESIII:2013fnz,CMS:2013fpt,ATLAS:2016kwu,BESIII:2019qvy,D0:2020nce,LHCb:2020fvo,LHCb:2020sey,LHCb:2020coc,CMS:2020eiw,LHCb:2021ten}. 
The radiative production process of $e^+e^-\to\gamma X(3872)$ was first observed by the BESIII experiment~\cite{BESIII:2013fnz}. It was later confirmed that these $X(3872)$ signals are produced through the radiative decay of a vector state, either $\psi(4160)$ or $Y(4230)$~\cite{BESIII:2019qvy}. In 2023, the BESIII experiment reported a new production process $e^+e^-\to\omega X(3872)$~\cite{BESIII:2022bse}, its cross section showing an enhancement around 4.75 GeV, which might be evidence for a potential resonance. It is therefore natural to search for the radiative process $Y\to\gamma X(3872)$ at this energy.
Moreover, the $X(3872)$ is considered a possible candidate of a $P$-wave charmonium $\chi_{c1}(2P)$ or containing a sizeable component of $\chi_{c1}(2P)$~\cite{Meng:2005er,Matheus:2009vq,Coito:2012vf}.  
By comparing the production of $X(3872)$ with that of $P$-wave charmonium states, $\chi_{c1}(1P)$ and $\chi_{c2}(1P)$, we can obtain more information about the production dynamics of the $X(3872)$ and hence its internal components.

In this article, we present a study of the processes $e^+e^-\to\omega X(3872)$ and $e^+e^-\to\gamma X(3872)$, using data samples with an integrated luminosity of $4.5~\text{fb}^{-1}$ taken at center-of-mass (c.m.) energies ranging from 4.66 to 4.95~GeV~\cite{BESIII:2022ulv}. We measure the branching fraction ratio $R$ through the $e^+e^-\to\omega X(3872)$ production channel. 
Additionally, we compare the $X(3872)$ and $\chi_{cJ}(1P)$ ($J=1,~2$) states by measuring the ratio of the cross sections of $e^+e^-\to\omega X(3872)$ to $e^+e^-\to\omega\chi_{cJ}(1P)$, $\sigma_{\omega X(3872)}/\sigma_{\omega\chi_{cJ}}$.
Furthermore, we search for the process of $e^+e^-\to\gamma X(3872)$ at c.m.~energies from 4.66 to 4.95 GeV, and measure the ratio of the cross sections of $e^+e^-\to\gamma X(3872)$ to $e^+e^-\to\omega X(3872)$, $\sigma_{\gamma X(3872)}/\sigma_{\omega X(3872)}$. 
The $\omega$ and $\chi_{cJ}$ candidates are reconstructed through the final states of $\pi^+\pi^-\pi^0$ and $\gamma J/\psi$, respectively, where $J/\psi$ meson decays to $\ell^+\ell^-$ ($\ell=e,~\mu$).
The decay modes $X(3872)\to\pi^+\pi^- J/\psi$ and $X(3872)\to\gamma J/\psi$ are both taken into account.

\section{BESIII detector and MC sample}
The BESIII detector~\cite{BESIII:2009fln} records symmetric $e^+e^-$ collisions 
provided by the BEPCII storage ring~\cite{Yu:2016cof}
in the center-of-mass energy range from 2.0 to 4.95~GeV,
with a peak luminosity of $1 \times 10^{33}\;\text{cm}^{-2}\text{s}^{-1}$ 
achieved at $\sqrt{s} = 3.77\;\text{GeV}$. 
BESIII has collected large data samples in this energy region~\cite{BESIII:2020nme,Lu2020Sep,Zhang:2022bdc}. The cylindrical core of the \mbox{BESIII} detector covers 93\% of the full solid angle and consists of a helium-based
multilayer drift chamber~(MDC), a plastic scintillator time-of-flight
system~(TOF), and a CsI(Tl) electromagnetic calorimeter~(EMC),
which are all enclosed in a superconducting solenoidal magnet
providing a 1.0~T magnetic field. The solenoid is supported by an octagonal flux-return yoke with resistive plate counter muon
identification modules (MUC) interleaved with steel. 
The charged-particle momentum resolution at $1~{\rm GeV}/c$ is
$0.5\%$, and the 
${\rm d}E/{\rm d}x$
resolution is $6\%$ for electrons
from Bhabha scattering. The EMC measures photon energies with a
resolution of $2.5\%$ ($5\%$) at $1$~GeV in the barrel (end cap)
region. The time resolution in the TOF barrel region is 68~ps, while
that in the end cap region is 60~ps~\cite{Li2017Dec,Guo2017Dec,Cao:2020ibk}. 

Simulated samples produced with a {\sc geant4}-based~\cite{GEANT4:2002zbu} Monte
Carlo (MC) software, which includes the geometric description of the
BESIII detector and the detector response, are used to determine
detection efficiencies and to estimate backgrounds. The signal MC
samples of $e^+e^-\to\omega X(3872)$ and
$e^+e^-\to\gamma X(3872)$ are simulated at each c.m.~energy, with $\omega\to\pi^+\pi^-\pi^0$, $X(3872)\to\pi^+\pi^-J/\psi$, $\gamma J/\psi$, and $J/\psi\to e^+e^-$, $\mu^+\mu^-$ being simulated according to the
branching fractions taken from the Particle Data Group
(PDG)~\cite{pdg}. The normalized channel $e^+e^-\to\omega\chi_{cJ}$, with $\chi_{cJ}\to\gamma J/\psi$ is also simulated at each c.m.~energy. The simulation models the beam energy
spread and initial state radiation (ISR) in the $e^+e^-$ annihilations
with the generator {\sc kkmc}~\cite{ref:kkmc1,ref:kkmc2}.  
The inclusive MC sample, with a luminosity that is ten times larger than the data, includes the production of open charm processes, the ISR production of vector charmonium(-like) states, and the continuum processes simulated with {\sc kkmc}~\cite{ref:kkmc1,ref:kkmc2}. 
All particle decays are modelled with {\sc
evtgen}~\cite{ref:evtgen1,ref:evtgen2} using branching fractions either taken from the PDG~\cite{pdg}, when available, or otherwise estimated with {\sc lundcharm}~\cite{ref:lundcharm1,ref:lundcharm2}. Final state radiation~(FSR) from charged final state particles is incorporated using the {\sc photos} package~\cite{photos}.

\section{Event Selection}

Charged tracks detected in the MDC are required to be within a polar
angle ($\theta$) range of $|\cos\theta|<0.93$ (the coverage of the MDC), where $\theta$ is
defined with respect to the $z$-axis, which is the symmetry axis of
the MDC. The
distance of closest approach to the interaction point (IP) must be
less than 10\,cm along the $z$-axis, $|V_{z}|<10~\text{cm}$, and less than
1\,cm in the transverse plane, $|V_{xy}|<1~\text{cm}$.

Photon candidates are identified using showers in the EMC.  The
deposited energy of each shower must be greater than 25~MeV in the
barrel region ($|\cos \theta|< 0.80$) and greater than 50~MeV in the
end cap region ($0.86 <|\cos \theta|< 0.92$). To exclude the showers
that originate from charged tracks, the angle between the position of
each shower in the EMC and the closest extrapolated charged track must
be greater than 10 degrees. To suppress the electronic noise and the showers
unrelated to the event, the difference between the EMC time (read out from the seed crystal with maximum deposited energy) and the
event start time is required to be within [0, 700]\,ns.

For the analyzed processes, the charged pions from $\omega$ or $X(3872)$ and the leptons from $J/\psi$ can be effectively distinguished by their momenta in the lab-frame. Charged tracks with momentum less or greater than $1~\text{GeV}/c$ are assigned as pions or leptons, respectively. The amount of deposited energy of the lepton candidates in the EMC is further used to separate muons from electrons. For muon candidates, the deposited energy in the EMC is required to be less than 0.4~GeV, while it is greater than 1.0~GeV for electrons.

\subsection{$\omega\gamma J/\psi$ events \label{sec:wgj}}
For the candidate events of $\omega X(3872)$ with $X(3872)\to\gamma J/\psi$, and $\omega\chi_{cJ}$ with $\chi_{cJ}\to \gamma J/\psi$, four charged tracks ($\pi^+\pi^-\ell^+\ell^-$) and at least two of the three photons in the final state are required to be reconstructed. One of the photon candidates is always ignored and not required to be detected in order to improve the selection efficiency.
To improve the resolution and suppress the background, a one-constraint (1C) kinematic fit is performed with a $e^+e^-\to\pi^+\pi^-\gamma_1\gamma_2\gamma_{\text{miss}}\ell^+\ell^-$ hypothesis, where the mass of the missing particle is constrained to zero ($\sqrt{(P_{e^+e^-}-P_{\pi^+\pi^-\gamma_1\gamma_2\ell^+\ell^-})^2}=0$, with $P$ denoting the four-momentum). If there is more than one combination of photons satisfying the kinematic constraint in an event, the one with the minimum $\chi^2$ from the kinematic fit is retained. To distinguish between the photons originating from the $\pi^0$ and  the $X(3872)/\chi_{cJ}$, a kinematic fit constraining also the invariant mass (2C) of two photons to the $\pi^0$ nominal mass  is applied. The combination with the minimum $\chi^2$ from the 2C kinematic fit is assigned as the right combination, and events with $\chi^2<20$ are selected.

To veto the background events from $e^+e^-\to\pi^+\pi^-\psi(3686)$, $\psi(3686)\to\pi^0\pi^0 J/\psi$, $\gamma\chi_{cJ}$, and $e^+e^-\to\pi^0\pi^0\psi(3686)$, $\psi(3686)\to\pi^+\pi^- J/\psi$, $|RM(\pi^+\pi^-)-m_{\psi(3686)}|>0.03~\text{GeV}/c^2$ and $|M(\pi^+\pi^- J/\psi)-m_{\psi(3686)}|>0.01~\text{GeV}/c^2$ are required, where $RM(\pi^+\pi^-)=\sqrt{(P_{e^+e^-}-P_{\pi^+\pi^-})^2}$ is the recoiling mass of $\pi^+\pi^-$ against the initial $e^+e^-$ collision system, $m_{\psi(3686)}$ is the $\psi(3686)$ nominal mass~\cite{pdg}, and $M(\pi^+\pi^- J/\psi)\equiv M(\pi^+\pi^-\ell^+\ell^-)-M(\ell^+\ell^-)+m_{J/\psi}$ is applied to improve resolution, where $m_{J/\psi}$ is the $J/\psi$ nominal mass~\cite{pdg}.
To reduce the $\mu/\pi$ misidentification background in the $J/\psi\to \mu^+\mu^-$ channel, e.g. $e^+e^-\to\omega\pi^+\pi^-$ events, at least one of the muon candidates must have a hit depth larger than $30$~cm in the MUC.

\begin{figure*}[htbp]
    \centering
    \includegraphics[width=0.48\linewidth]{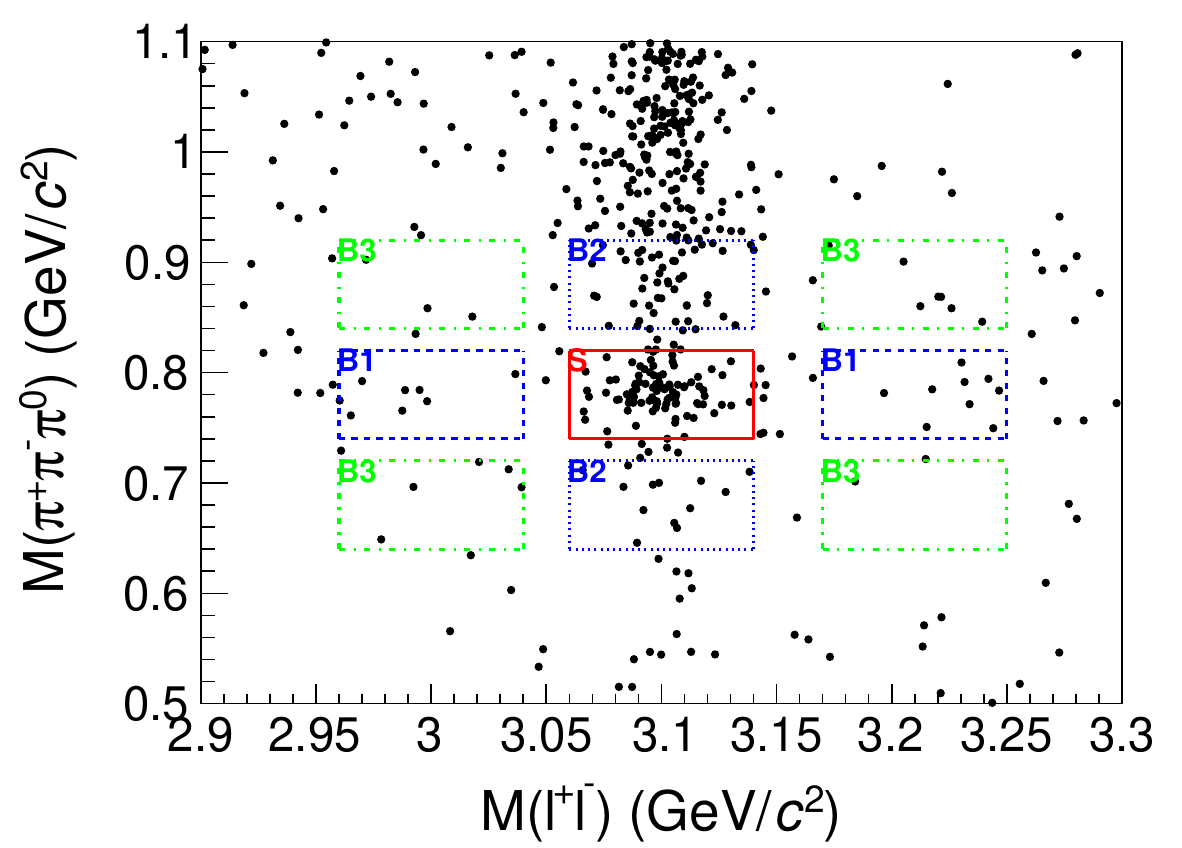}
    \includegraphics[width=0.48\linewidth]{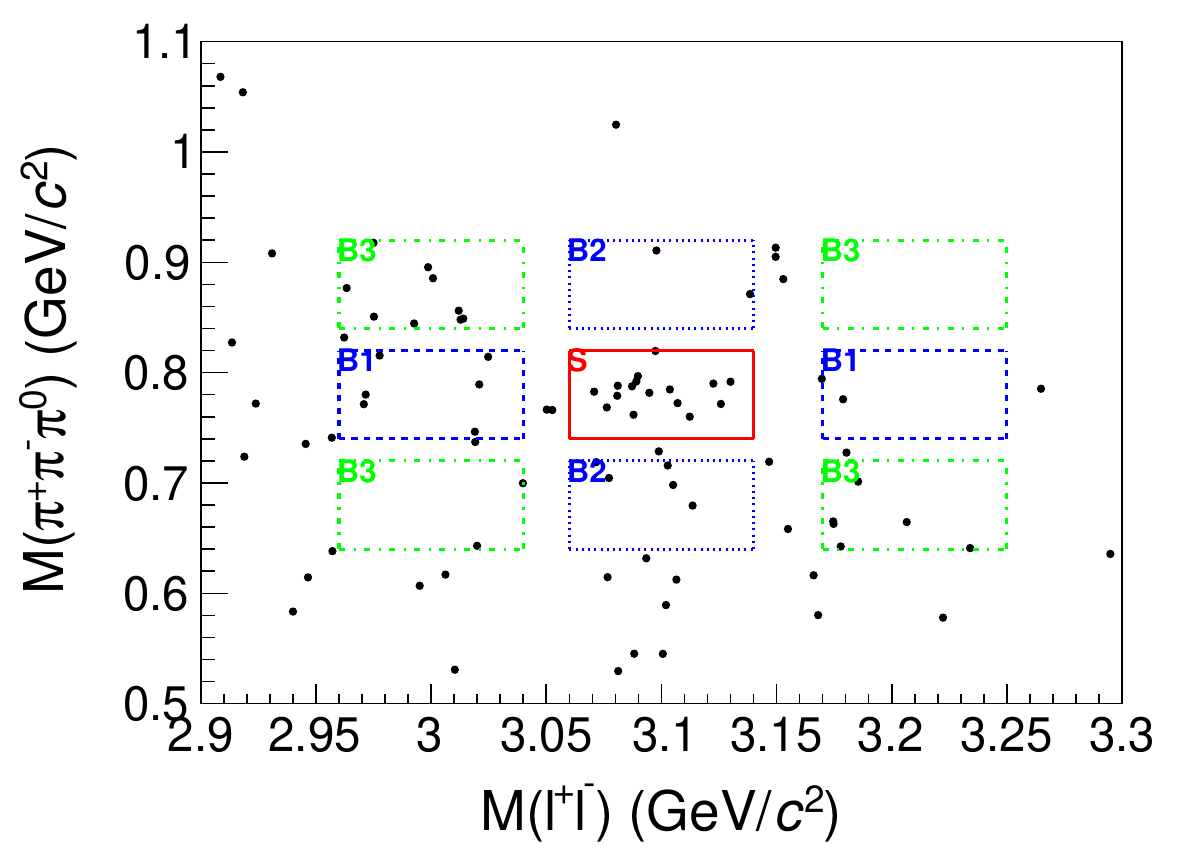}
    \caption{The 2D distributions of $M(\pi^+\pi^-\pi^0)$ versus $M(\ell^+\ell^-)$ with $M(\gamma J/\psi)\in [3.45,~3.62]~\text{GeV}/c^2$ (left) and $[3.82,~3.92]~\text{GeV}/c^2$ (right). The dots are data samples, the red solid boxes are $\omega J/\psi$ signal region (S), the blue dashed, blue dotted and green dash-dotted boxes indicate $\omega$ non-$J/\psi$ (B1), $J/\psi$ non-$\omega$ (B2) and non-$\omega$ non-$J/\psi$ (B3) sideband regions, respectively.
    }
    \label{fig:w-j}
\end{figure*}

Figure~\ref{fig:w-j} shows the two-dimensional (2D) distribution of $M(\pi^+\pi^-\pi^0)$ versus $M(\ell^+\ell^-)$ after the aforementioned  requirements have been applied. The $\omega$ and $J/\psi$ signals are observed in the data with $M(\gamma J/\psi)\in [3.45,~3.62]~\text{GeV}/c^2$ and $[3.82,~3.92]~\text{GeV}/c^2$ corresponding to the $\chi_{cJ}$ and $X(3872)$ signal regions. The $\omega$ and $J/\psi$ signal regions are defined as $M(\pi^+\pi^-\pi^0)\in[0.74,~0.82]~\text{GeV}/c^2$ and $M(\ell^+\ell^-)\in[3.06,~3.14]~\text{GeV}/c^2$, respectively. The $\omega$ and $J/\psi$ sidebands are used to estimate the non-$\omega$ and non-$J/\psi$ backgrounds. The $\omega$ sideband region is defined as $M(\pi^+\pi^-\pi^0)\in[0.64,~0.72]\cup[0.84,~0.92]~\text{GeV}/c^2$, which is twice as wide as the $\omega$ signal region. The $J/\psi$ sideband region is defined as $M(\ell^+\ell^-)\in[2.96,~3.04]\cup[3.17,~3.25]~\text{GeV}/c^2$, which is twice as wide as the $J/\psi$ signal region.  

\begin{figure}[htbp]
    \centering
    \includegraphics[width=\linewidth]{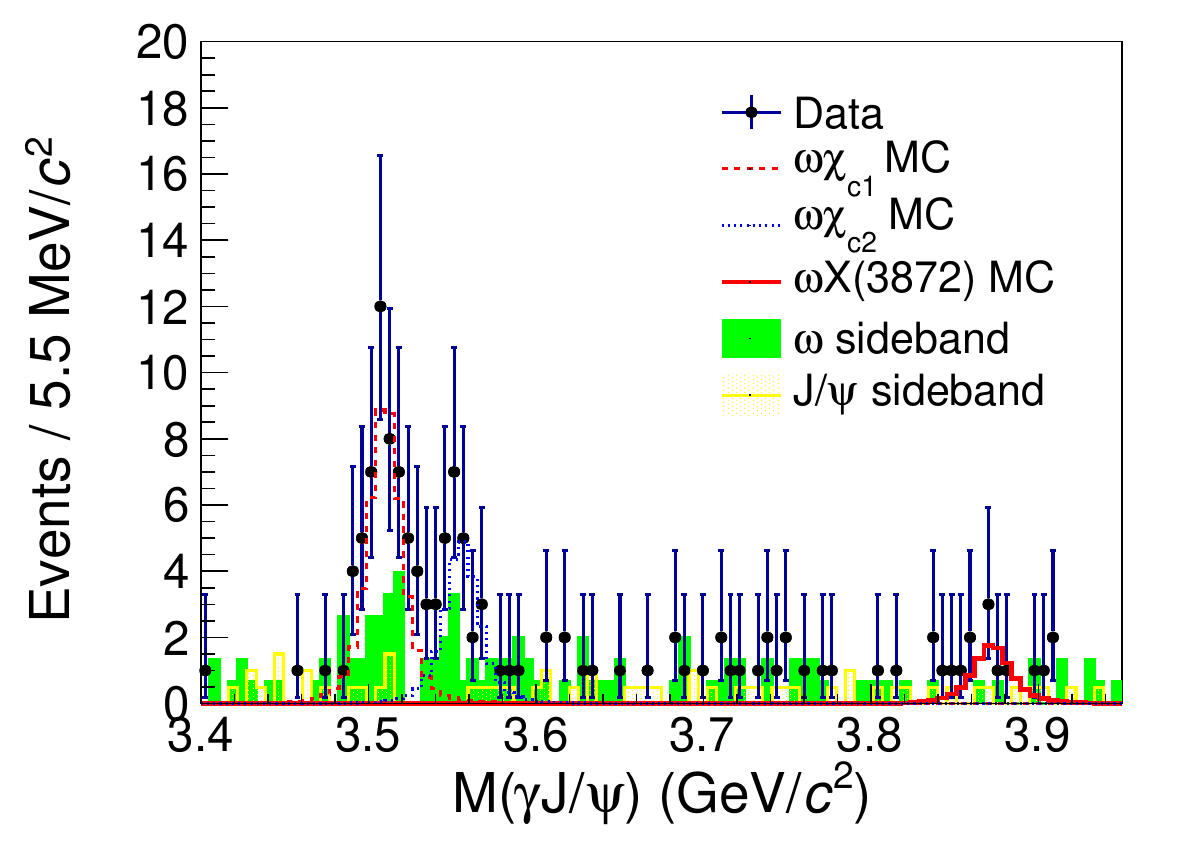}
    \caption{The distribution of $M(\gamma J/\psi)$. The dots with error bars are data samples, the red dashed, blue dotted and red solid histograms are $\omega\chi_{c1}$, $\omega\chi_{c2}$ and $\omega X(3872)$ MC, respectively. The green filled and yellow shaded histograms represent $\omega$ and $J/\psi$ sidebands, respectively.}
    \label{fig:mgj}
\end{figure}

Figure~\ref{fig:mgj} displays the invariant mass distribution of $\gamma J/\psi$ within the $\omega$ and $J/\psi$ mass windows. The $\chi_{c1}$ and $\chi_{c2}$ signals are significant, and there are several events near $3.872~\text{GeV}/c^2$ that correspond to the $X(3872)$. The $\chi_{cJ}$ peaking background from the $\omega$ sideband is also evident, which could be attributed to $\pi^+\pi^-\pi^0\chi_{cJ}$ or $\omega^*\chi_{cJ}$ events. Here, the $\omega^*$ represents a higher excited state of $\omega$, such as $\omega(1420)$.

\subsection{$\omega\pi^+\pi^- J/\psi$ events \label{sec:wppj}}

The event selection criteria for $\omega X(3872)$ with $X(3872)\to\pi^+\pi^- J/\psi$ follow Ref.~\cite{BESIII:2022bse}. Candidate events with five charged tracks and two photons ($\pi^+\pi^-\pi^\pm\ell^+\ell^-\gamma\gamma$) are reconstructed and referred to as 5-track events. Additionally, to improve the selection efficiency, candidate events with six charged tracks and at least one photon ($\pi^+\pi^-\pi^+\pi^-\ell^+\ell^-\gamma$) are also reconstructed and referred to as 6-track events. 

\begin{figure*}[htbp]
    \centering
    \includegraphics[width=0.48\linewidth]{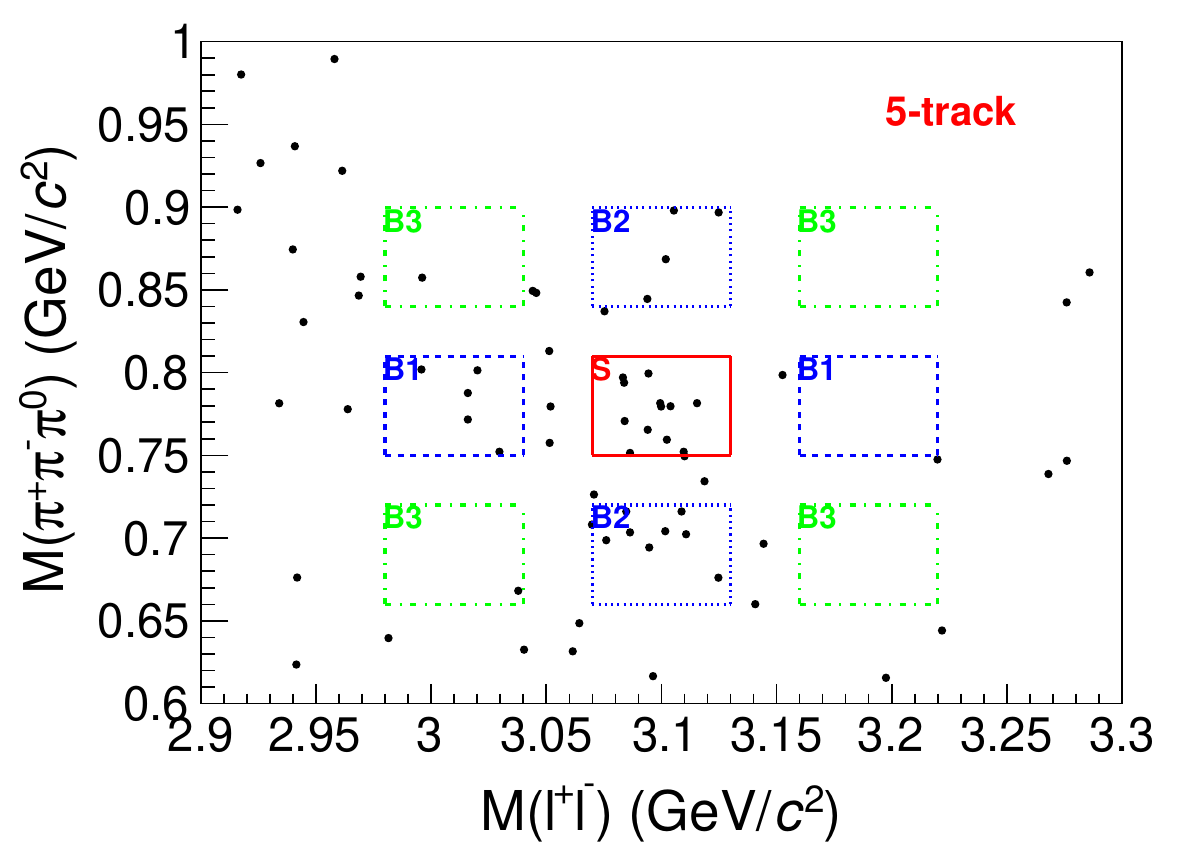}
    \includegraphics[width=0.48\linewidth]{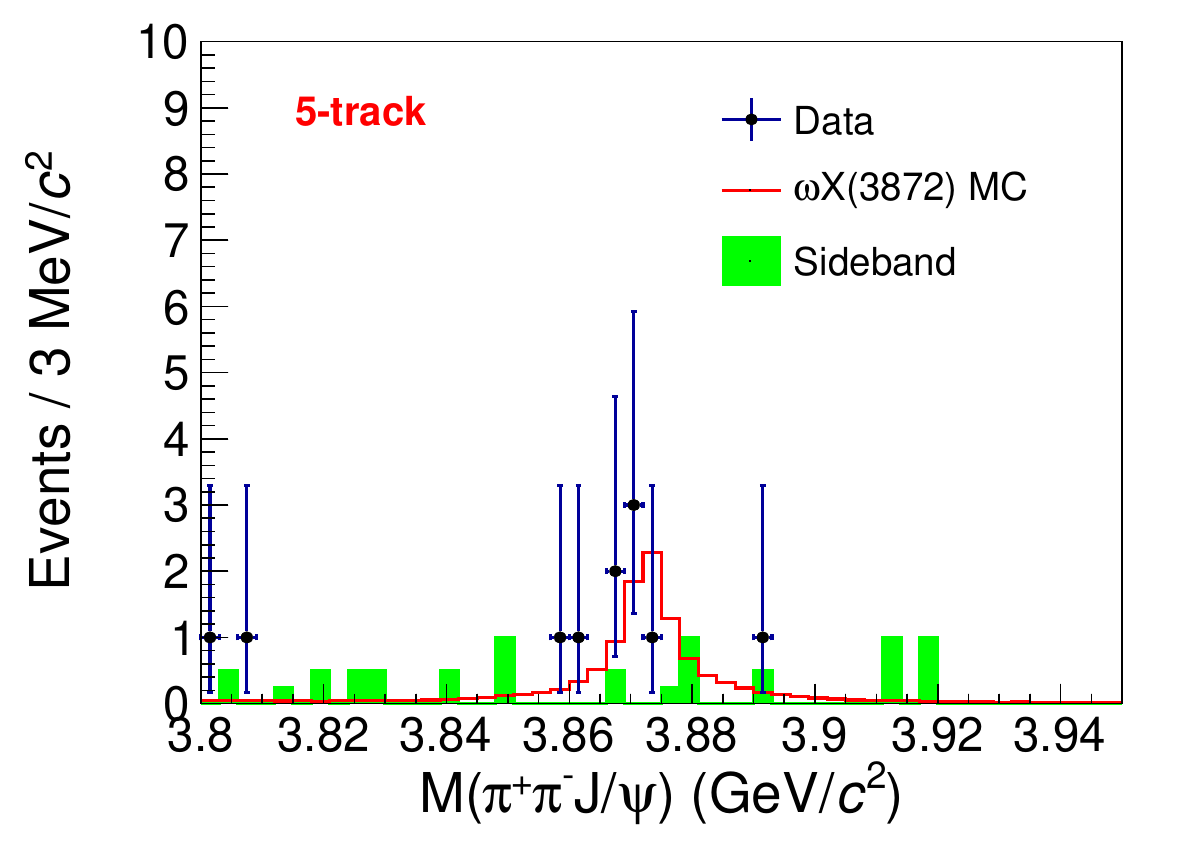}
    \includegraphics[width=0.48\linewidth]{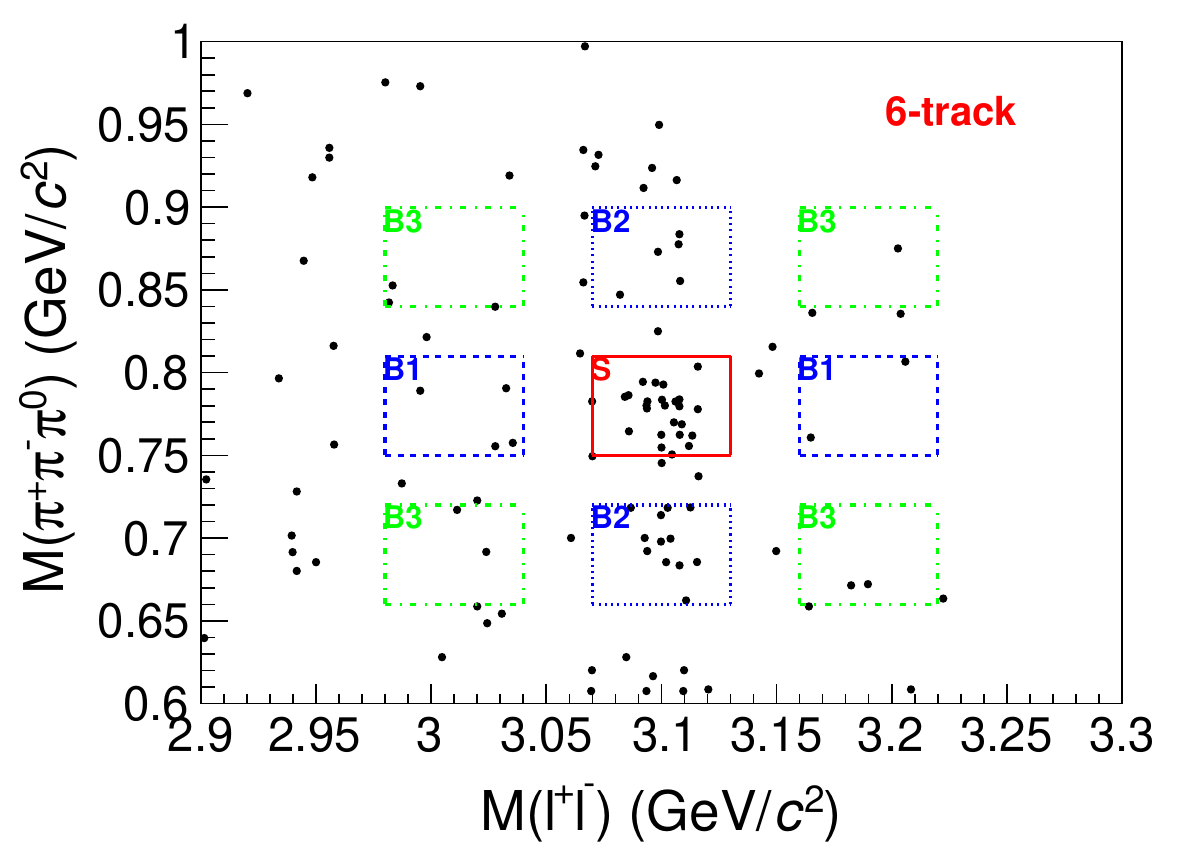}
    \includegraphics[width=0.48\linewidth]{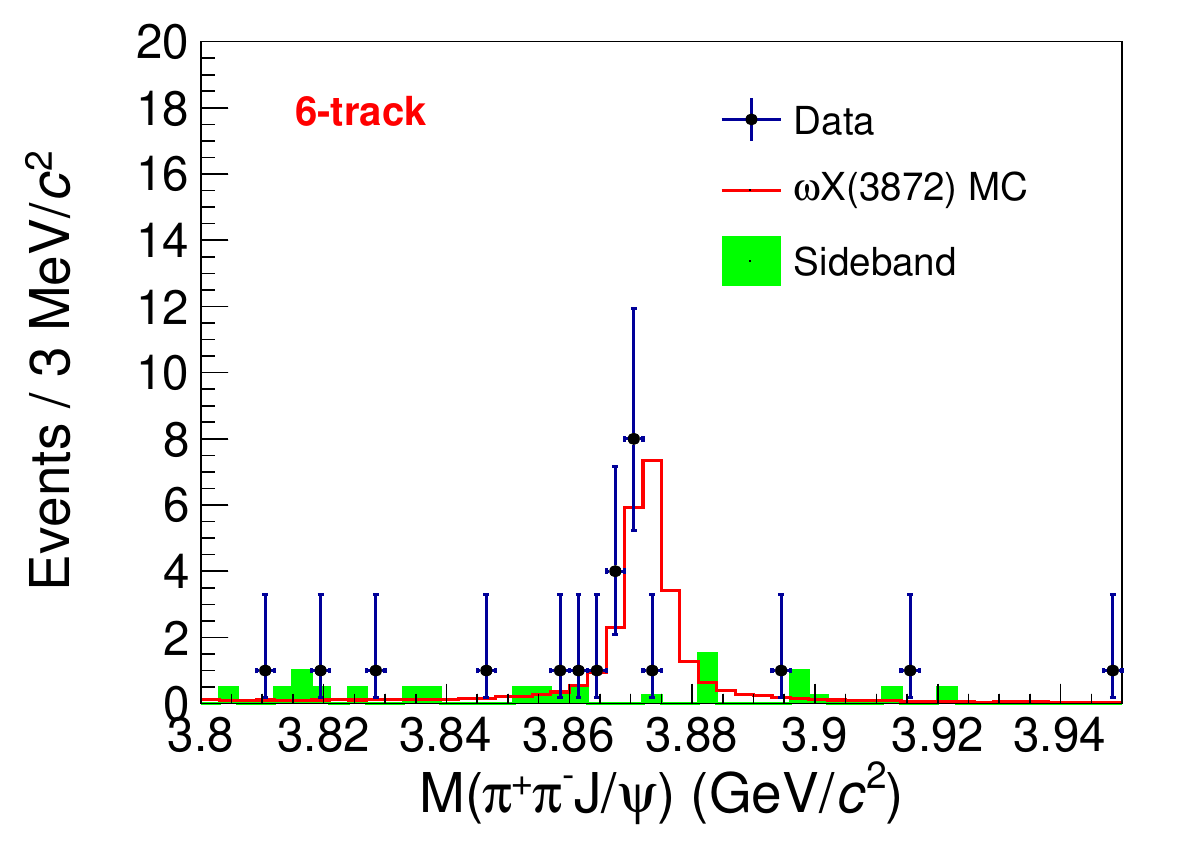}
    \caption{The 2D distributions of $M(\pi^+\pi^-\pi^0)$ versus $M(\ell^+\ell^-)$ (left column) and the distributions of $M(\pi^+\pi^- J/\psi)$ (right column) for the 5-track (upper row) and 6-track (bottom row) events of  $\omega\pi^+\pi^- J/\psi$. In the left column, the dots are data samples, the red solid boxes are $\omega J/\psi$ signal region (S), the blue dashed, blue dotted and green dash-dotted boxes indicate $\omega$ non-$J/\psi$ (B1), $J/\psi$ non-$\omega$ (B2) and non-$\omega$ non-$J/\psi$ (B3) sideband regions, respectively. In the right column, the dots with error bars are data samples, the red solid histograms are $\omega X(3872)$ MC, and the green filled histograms represent $\omega$-$J/\psi$ 2D sidebands (B1/2+B2/2-B3/4).}
    \label{fig:wppj}
\end{figure*}

Figure~\ref{fig:wppj} shows the 2D distributions of $M(\pi^+\pi^-\pi^0)$ versus $M(\ell^+\ell^-)$, and the distributions of $M(\pi^+\pi^- J/\psi)$ for the 5-track and 6-track events. In both cases, the $\omega$ and $J/\psi$ signal regions are defined as $M(\pi^+\pi^-\pi^0)\in[0.75,~0.81]~\text{GeV}/c^2$ and $M(\ell^+\ell^-)\in[3.07,~3.13]~\text{GeV}/c^2$, respectively. The $\omega$ and $J/\psi$ sideband regions are defined as $M(\pi^+\pi^-\pi^0)\in[0.66,~0.72]\cup[0.84,~0.90]~\text{GeV}/c^2$ and $M(\ell^+\ell^-)\in[2.98,~3.04]\cup[3.17,~3.23]~\text{GeV}/c^2$, respectively. The defined $\omega$ and $J/\psi$ sideband regions are twice as wide as their respective signal regions.

\subsection{$\gamma\pi^+\pi^- J/\psi$ events}

The event selection criteria for $\gamma X(3872)$ with $X(3872)\to\pi^+\pi^- J/\psi$ follow Ref.~\cite{BESIII:2013fnz}. Candidate events with four charged tracks and at least one photon ($\pi^+\pi^-\ell^+\ell^-\gamma$) are reconstructed. 

\begin{figure}[htbp]
    \centering
    \includegraphics[width=\linewidth]{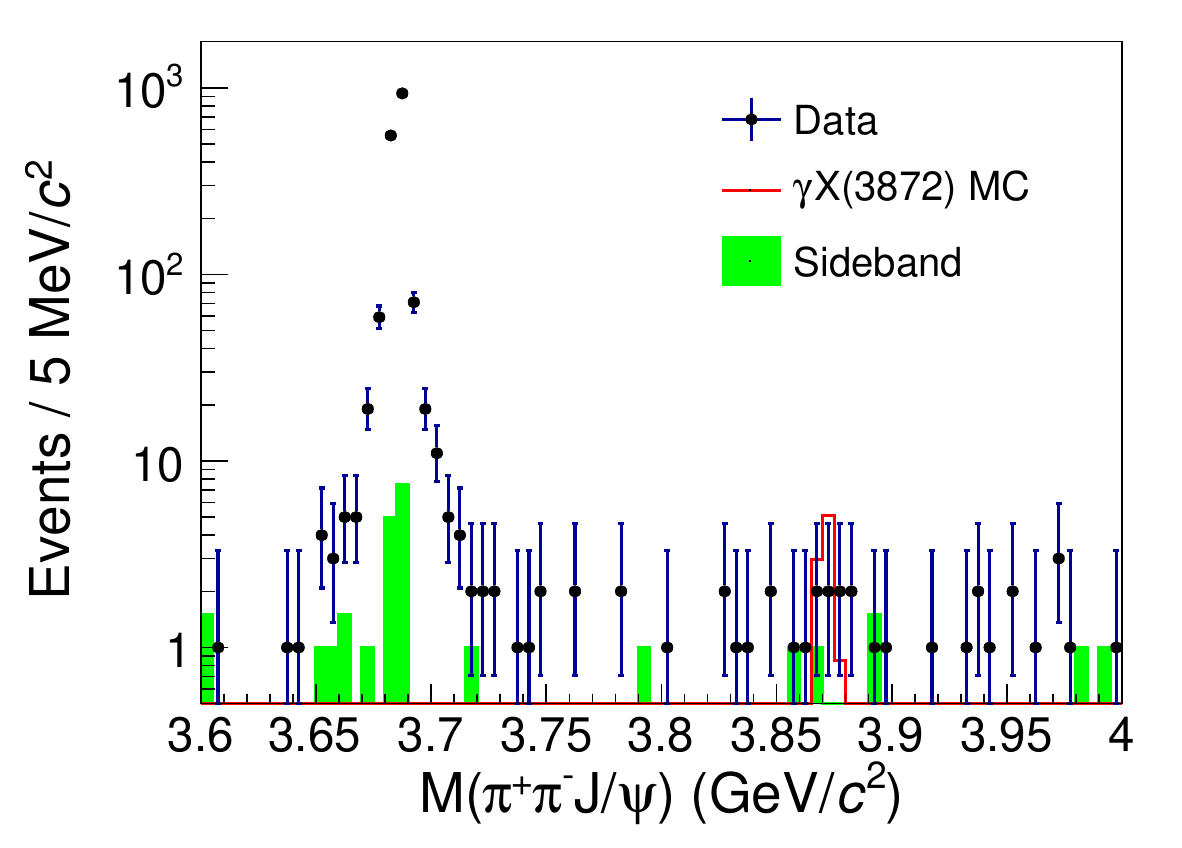}
    \caption{The distribution of $M(\pi^+\pi^- J/\psi)$. The dots with error bars are data samples, the red solid and green filled histograms are $\gamma X(3872)$ MC and $J/\psi$ sideband, respectively.  }
    \label{fig:gx}
\end{figure}

The $J/\psi$ signal region is defined as $M(\ell^+\ell^-)\in[3.08,~3.12]~\text{GeV}/c^2$, and the sideband region is $M(\ell^+\ell^-)\in[3.02,~3.06]\cup[3.14,~3.18]~\text{GeV}/c^2$ which is twice as wide as the signal region. After applying the selection criteria for $\gamma X(3872)$ events, the invariant mass distribution of $\pi^+\pi^- J/\psi$ is displayed in Fig.~\ref{fig:gx}, where the $\psi(3686)$ peak corresponding to the $e^+e^-\to\gamma_\text{ISR}\psi(3686)$ events is significant, but no obvious $X(3872)$ signal is observed in the data.

\section{$R$ measurement\label{sec:R}}

\begin{figure*}[htbp]
	\centering
	\includegraphics[width=\linewidth]{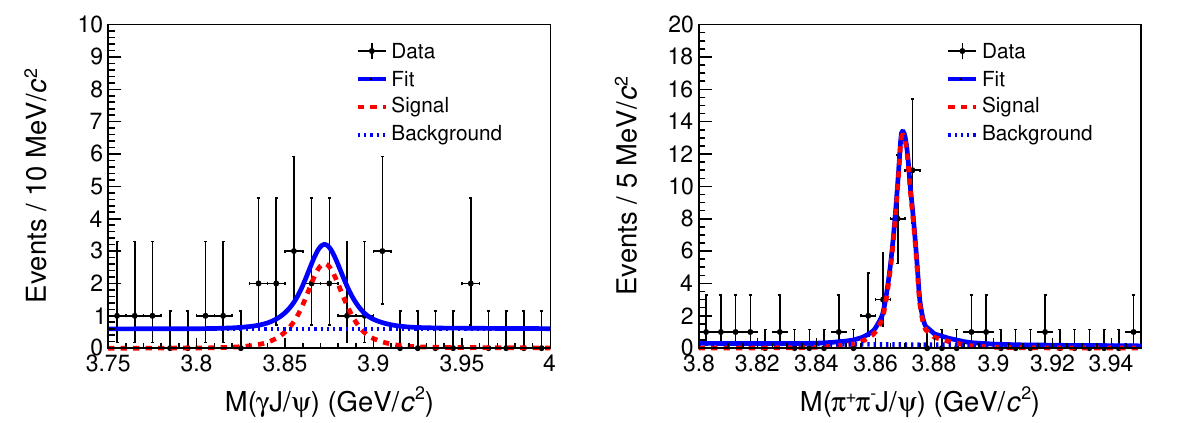}
	\caption{The unbinned maximum likelihood fits to the  distributions of $M(\gamma J/\psi)$ (left) and $M(\pi^+\pi^- J/\psi)$ (right). The dots with error bars are data samples, the blue solid curves are the fit results, the red dashed and blue dotted lines represent signal and background shapes, respectively. \label{fig:simfit-wx}}
\end{figure*}

The ratio of the branching fraction of $X(3872)\to\gamma J/\psi$ to that of $X(3872)\to\pi^+\pi^- J/\psi$ is calculated by 
\begin{linenomath*}
\begin{equation}\label{eq:R}
	R=\frac{N_{\gamma J/\psi}}{N_{\pi^+\pi^- J/\psi}}
 \frac{\varepsilon^{\text{Ave.}}_{\pi^+\pi^- J/\psi}}{\varepsilon^{\text{Ave.}}_{\gamma J/\psi}},
\end{equation} 
\end{linenomath*}
where $N_{\gamma J/\psi}$ and $N_{\pi^+\pi^- J/\psi}$ are the numbers of signal events for $X(3872)\to\gamma J/\psi$ and $X(3872)\to\pi^+\pi^- J/\psi$ modes of the $e^+e^-\to \omega X(3872)$ process, respectively. $\varepsilon^{\text{Ave.}}_{\gamma J/\psi}$ and $\varepsilon^{\text{Ave.}}_{\pi^+\pi^- J/\psi}$ are the average selection efficiencies at c.m.~energies between 4.66 and 4.95 GeV for $X(3872)\to\gamma J/\psi$ and $X(3872)\to\pi^+\pi^- J/\psi$, respectively. The average selection efficiency is defined as $\varepsilon^{\text{Ave.}}=\sum_{i}\mathcal{L}_i\sigma_i\varepsilon_i/\sum_{i}\mathcal{L}_i\sigma_i$, where $\mathcal{L}_i$, $\sigma_i$, and $\varepsilon_i$ are the luminosity, cross section, and selection efficiency at the $i$th c.m.~energy.
To determine the number of signal events, unbinned maximum likelihood fits are performed to the distributions of $M(\gamma J/\psi)$ in the $X(3872)\to\gamma J/\psi$ mode and $M(\pi^+\pi^- J/\psi)$ in the $X(3872)\to\pi^+\pi^- J/\psi$ mode, as shown in Fig.~\ref{fig:simfit-wx}. The signal probability-density-function (PDF) is described by an MC-simulated shape convolved with a Gaussian function, which models the resolution difference between data and MC simulation. The Gaussian parameters in the $M(\gamma J/\psi)$ fit are constrained to the values extracted from $\chi_{cJ}\to\gamma J/\psi$ events, which provide a higher statistical control sample. The Gaussian parameters are free in the $M(\pi^+\pi^- J/\psi)$ fit. In both cases, a linear function is used to model the background. The fit yields $N_{\gamma J/\psi}=8.1\pm3.9$ and $N_{\pi^+\pi^- J/\psi}=24.2\pm5.2$, where the uncertainties are statistical only.

Based on the fit results of Fig.~\ref{fig:simfit-wx} and Eq.~(\ref{eq:R}), the ratio $R=0.38\pm0.20$ is obtained.
The statistical significance of the $X(3872)\to\gamma J/\psi$ signal is estimated by comparing the difference in the log-likelihood values [$\Delta(-2\ln\mathcal{L})=5.1$] with and without the signal in the fit, and also taking the change of the number of degrees of freedom ($\Delta \text{ndf}=1$) into consideration. The statistical significance is found to be $2.3\sigma$, and an upper limit of $R$ ($R^\text{UL}$) is determined via a Bayesian approach~\cite{pdg}. A likelihood profile $L(R)$ is performed with various assumptions for the value of $R$ in the fit. 
To incorporate the systematic uncertainty into the upper limit, the likelihood distribution is convolved with a Gaussian function with a width equal to the systematic uncertainty (cf.~Section~\ref{sec:sys-1}). The $R^\text{UL}$ at 90\% confidence level (C.L.) is determined to be 0.83 by $\int_{0}^{R^{\text{UL}}} L(R) \mathrm{d} R / \int_{0}^{+\infty} L(R)\mathrm{d} R = 0.9$. For comparison, Table~\ref{tab:R} lists $R$ measured by BESIII with $e^+e^-\to\gamma X(3872)$ and Belle with $B^\pm\to K^\pm X(3872)$.

\begin{table}[htpb]
	\renewcommand\arraystretch{1.4}
	\centering
	\caption{The $R$ measured by BESIII with $e^+e^-\to\gamma X(3872)$, Belle with $B^\pm\to K^\pm X(3872)$, and this work, where the uncertainty includes both statistical and systematic. The last row shows the average of the three results.  \label{tab:R}}
	\begin{tabular}{cc}
		\hline
		\hline
		Experiment & $R$ \\
		\hline
		
		\makecell[c]{ 	BESIII~\cite{BESIII:2020nbj}  \\ $[e^+e^-\to\gamma X(3872)]$} & $0.79\pm0.28$ \\
		\hline
		
		\makecell[c]{Belle~\cite{Belle:2011vlx,Belle:2011wdj} \\ $[B^\pm\to K^\pm X(3872)]$} & $0.21\pm0.06$ \\
		\hline
		
		\makecell[c]{This work \\ $[e^+e^-\to\omega X(3872)]$} & 
  
		\makecell[c]{$0.38\pm0.20$ ($<0.83$) } \\

  		\hline
		Average & $0.25\pm0.06$ \\
		
		\hline
		\hline
	\end{tabular}
\end{table}

\section{$\sigma_{\omega X(3872)}/\sigma_{\omega\chi_{cJ}}$ measurement}

The ratio of the cross section of $e^+e^-\to\omega X(3872)$ to that of $e^+e^-\to\omega \chi_{cJ}$ is derived by
\begin{linenomath*}
\begin{equation}\label{eq:ratio-xcj}
\begin{split}
    \frac{\sigma_{\omega X(3872)}}{\sigma_{\omega\chi_{cJ}}} =
 &\frac{N_{\omega X(3872)}}{N_{\omega\chi_{cJ}}}
  \\
 &\times
 \frac{[\varepsilon(1+\delta)]_{\omega\chi_{cJ}}^\mathrm{Ave.}\mathcal{B}_{\chi_{cJ}}}
 {\{[\varepsilon(1+\delta)]_{\pi^+ \pi^- J/\psi}^\mathrm{Ave.}+[\varepsilon(1+\delta)]_{\gamma J/\psi}^\mathrm{Ave.}R\}\mathcal{B}_{X(3872)}},
\end{split}
\end{equation}
\end{linenomath*}
where $N_{\omega X(3872)}$ is the total signal yield of the $e^+e^-\to\omega X(3872)$ with $X(3872)\to \gamma J/\psi$ and $\pi^+\pi^- J/\psi$ modes, $N_{\omega\chi_{cJ}}$ is the signal yield of  $e^+e^-\to\omega\chi_{cJ}$ with $\chi_{cJ}\to \gamma J/\psi$. $[\varepsilon(1+\delta)]_{\omega\chi_{cJ}}^\mathrm{Ave.}$, $[\varepsilon(1+\delta)]_{\pi^+ \pi^- J/\psi}^\mathrm{Ave.}$ and $[\varepsilon(1+\delta)]_{\gamma J/\psi}^\mathrm{Ave.}$ are the average of the product of selection efficiencies and ISR factors at c.m.~energies from 4.66 to 4.95 GeV for these processes, defined as $[\varepsilon(1+\delta)]^{\text{Ave.}}=\sum_{i}\mathcal{L}_i\sigma_i\varepsilon_i(1+\delta)_i/\sum_{i}\mathcal{L}_i\sigma_i$, where $\mathcal{L}_i$, $\sigma_i$, $\varepsilon_i$, and $(1+\delta)_i$ are the luminosity, cross section, selection efficiency, and ISR factor at the $i$th c.m.~energy. The ISR factor is calculated by {\sc kkmc} with an accuracy of 0.1\%~\cite{ref:kkmc1,ref:kkmc2}. $\mathcal{B}_{\chi_{cJ}}$ and $\mathcal{B}_{X(3872)}$ are the branching fractions of $\chi_{cJ} \to \gamma J/\psi$ and $X(3872)\to \pi^+\pi^- J/\psi$, respectively.

\begin{figure*}[htbp]
	\centering
	\includegraphics[width=\linewidth]{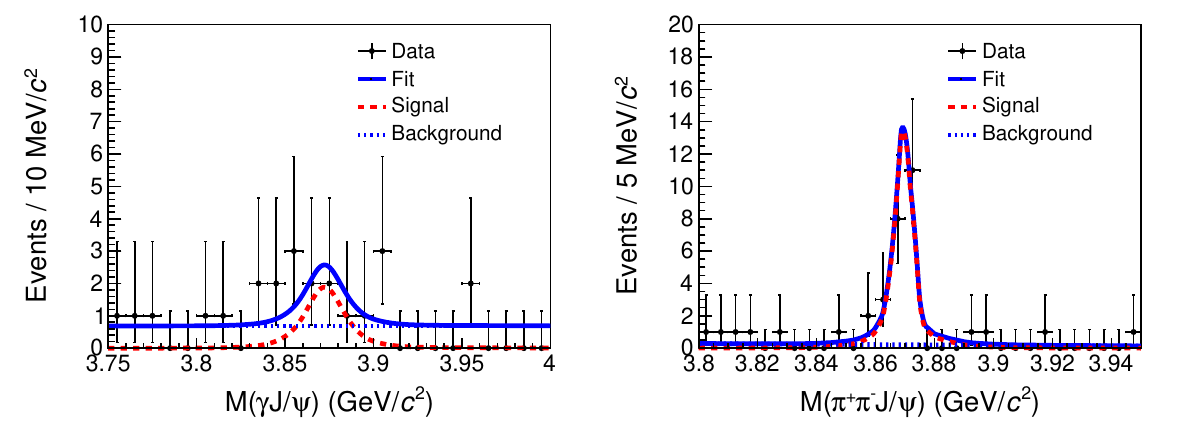}
	\caption{The simultaneous fit to the distributions of $M(\gamma J/\psi)$ (left) and $M(\pi^+\pi^- J/\psi)$ (right). The dots with error bars are data samples, the blue solid curves are the fit results, the red dashed and blue dotted lines represent signal and background shapes, respectively. \label{fig:simfit-wx-fixR}}
\end{figure*}

In the $\sigma_{\omega X(3872)}/\sigma_{\omega\chi_{cJ}}$ measurement, the two decay modes, $X(3872)\to\gamma J/\psi$ and $\pi^+\pi^- J/\psi$, are considered to extract the signal yields of $\omega X(3872)$. A simultaneous fit is conducted on the distributions of $M(\gamma J/\psi)$ in the $\gamma J/\psi$ mode and $M(\pi^+\pi^- J/\psi)$ in the $\pi^+\pi^- J/\psi$ mode, as shown in Fig.~\ref{fig:simfit-wx-fixR}. The PDF used in this fit is the same as that in the $R$ measurement (cf.~Fig.~\ref{fig:simfit-wx}), but $R$ is fixed to the average value of $0.25\pm0.06$. Taking into account the efficiency, ISR factor and branching fraction of $X(3872)\to\pi^+ \pi^- J/\psi$, the fit yields $\frac{N_{\omega X(3872)}}{\{[\varepsilon(1+\delta)]_{\pi^+\pi^- J/\psi}^\mathrm{Ave.}+[\varepsilon(1+\delta)]_{\gamma J/\psi}^\mathrm{Ave.}R\} \mathcal{B}_{X(3872)}}=3075\pm616$, which represents the number of produced $\omega X(3872)$ events. 

\begin{figure*}[htbp]
	\centering
	\includegraphics[width=\linewidth]{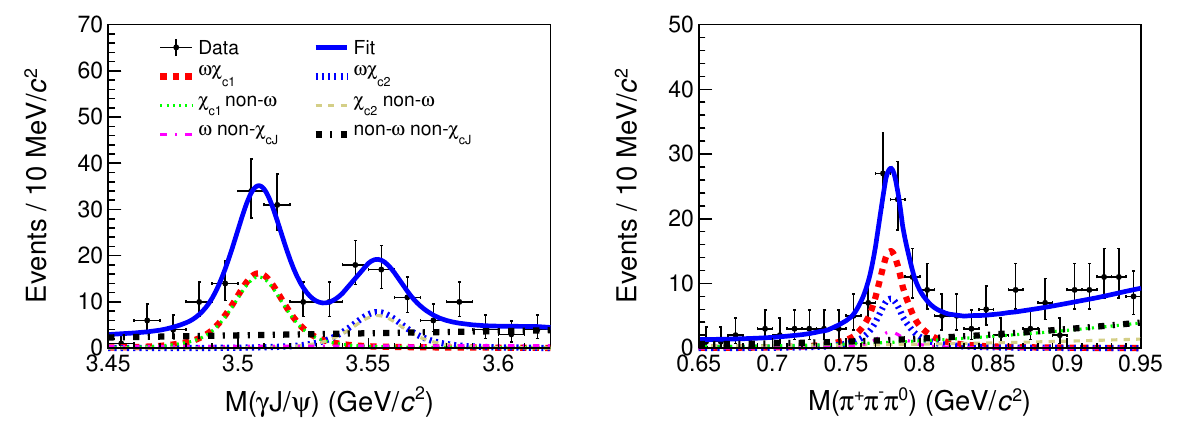}
	\caption{The 2D fit to the $M(\gamma J/\psi)$ and $M(\pi^+\pi^-\pi^0)$ distributions. The dots with error bars are data samples, the solid curves are the fit results, the thick dashed and dotted curves are $\omega\chi_{c1}$ ($S_{\omega}S_{\chi_{c1}}$) and $\omega\chi_{c2}$ ($S_{\omega}S_{\chi_{c2}}$) signals, respectively. The thin dotted, thin dashed, thin dash-dotted, and thick dash-dotted represent $\chi_{c1}$ non-$\omega$ ($B_{3\pi}S_{\chi_{c1}}$), $\chi_{c2}$ non-$\omega$ ($B_{3\pi}S_{\chi_{c2}}$), $\omega$ non-$\chi_{cJ}$ ($B_{\gamma J/\psi}S_{\omega}$), and non-$\omega$ non-$\chi_{cJ}$ ($B_{3\pi}B_{\gamma J/\psi}$), respectively. \label{fig:fit-wcj-2d}}
\end{figure*}

For the $\omega\chi_{cJ}$ events, there are potential peaking backgrounds from $\pi^+\pi^-\pi^0\chi_{cJ}$ or $\omega(1420)\chi_{cJ}$, which are observed in Fig.~\ref{fig:mgj}.
To extract the signal yield $N_{\omega \chi_{cJ}}$, a 2D fit is performed to the $M(\gamma J/\psi)$ and $M(\pi^+\pi^-\pi^0)$ distributions. The 2D PDF $f_\text{2D}$ is constructed with the following six components:
\begin{linenomath*}
\begin{equation}
	\begin{aligned}
	f_\text{2D} = &S_{\omega}S_{\chi_{c1}}+S_{\omega}S_{\chi_{c2}}\\
	&+B_{3\pi}S_{\chi_{c1}}+B_{3\pi}S_{\chi_{c2}}+B_{\gamma J/\psi}S_{\omega}+B_{3\pi}B_{\gamma J/\psi},
	\end{aligned}
\end{equation}
\end{linenomath*}
where $S_{\omega}S_{\chi_{cJ}}$, $B_{3\pi}S_{\chi_{cJ}}$, $B_{\gamma J/\psi}S_{\omega}$, and $B_{3\pi}B_{\gamma J/\psi}$ correspond to $\omega\chi_{cJ}$, $\chi_{cJ}$ non-$\omega$, $\omega$ non-$\chi_{cJ}$, and non-$\omega$ non-$\chi_{cJ}$ components, respectively. The symbols $S$ and $B$ stand for the signal shape and background shape, respectively. The signal shape is described by the MC-simulated shape convolved with a Gaussian resolution. The background shapes of $B_{\gamma J/\psi}$ and $B_{3\pi}$ are described by first-order and second-order polynomial functions, respectively. Figure~\ref{fig:fit-wcj-2d} shows the fit result, which yields $N_{\omega\chi_{c1}}=46.3\pm8.6$ and $N_{\omega\chi_{c2}}=23.5\pm7.2$, where the uncertainties are statistical. 
The significance of the $\chi_{cJ}$ signal is estimated using the same method mentioned above. With the changes of log-likelihood values $\Delta(-2\ln\mathcal{L})=47.0$ and 16.4 for the $\chi_{c1}$ and $\chi_{c2}$, respectively, as well as a change in the number of degrees of freedom $\Delta \text{ndf} =1$, the significance of $\omega\chi_{c1}$ and $\omega\chi_{c2}$ signals is estimated to be $6.8\sigma$ and $4.0\sigma$, respectively.

With the signal yields of $\omega X(3872)$ and $\omega\chi_{cJ}$, the cross section ratios of $\sigma_{\omega X(3872)}/\sigma_{\omega\chi_{c1}}=5.2\pm1.0$ and $\sigma_{\omega X(3872)}/\sigma_{\omega\chi_{c2}}=5.5\pm1.1$ are obtained from Eq.~(\ref{eq:ratio-xcj}), where the uncertainties are statistical only. The measured results and related parameters in Eq.~(\ref{eq:ratio-xcj}) are summarized in Table~\ref{tab:summary}.

\begin{table}[htbp]
	\renewcommand\arraystretch{1.5}
	\centering
	\caption{The results of $\sigma_{\omega X(3872)}/\sigma_{\omega\chi_{cJ}}$ and the values of related parameters in Eq.~(\ref{eq:ratio-xcj}). The first uncertainties are statistical, and the second systematic. 
 \label{tab:summary}}
	\begin{tabular}{cc}
		\hline
		\hline
Parameter & Value \\
		\hline
		\specialrule{0em}{0pt}{3pt}

		$\frac{N_{\omega X(3872)}}{\{[\varepsilon(1+\delta)]_{\pi^+\pi^- J/\psi}^\mathrm{Ave.}+[\varepsilon(1+\delta)]_{\gamma J/\psi}^\mathrm{Ave.}R\} \mathcal{B}_{X(3872)}}$
		 & $3075\pm616$   \\

		$N_{\omega\chi_{c1}}$ & $46.3\pm8.6$ \\
        $N_{\omega\chi_{c2}}$ & $23.5\pm7.2$ \\
		
	    $[\varepsilon(1+\delta)]_{\pi^+\pi^- J/\psi}^\mathrm{Ave.}$ & 0.22 \\
        $[\varepsilon(1+\delta)]_{\gamma J/\psi}^\mathrm{Ave.}$ & 0.19 \\

    $[\varepsilon(1+\delta)]_{\omega\chi_{c1}}^\mathrm{Ave.}$ & 0.23  \\
    $[\varepsilon(1+\delta)]_{\omega\chi_{c2}}^\mathrm{Ave.}$ & 0.22 \\

		$\mathcal{B}_{\chi_{c1}}$ & $(34.3\pm1.0)\%$~\cite{pdg} \\
        $\mathcal{B}_{\chi_{c2}}$ & $(19.0\pm0.5)\%$~\cite{pdg} \\
		$\mathcal{B}_{X(3872)}$ & $(3.8\pm 1.2)\%$~\cite{pdg} \\
        $R$ & $0.25\pm0.06$ \\

		\hline

		$\sigma_{\omega X(3872)}/\sigma_{\omega\chi_{c1}}$ & $5.2\pm1.0\pm1.9$  \\
		$\sigma_{\omega X(3872)}/\sigma_{\omega\chi_{c2}}$ & $5.5\pm1.1\pm2.4$ \\

		\hline
		\hline
	\end{tabular}
\end{table}

\section{$\sigma_{\gamma X(3872)}/\sigma_{\omega X(3872)}$ measurement}

The ratio of the cross sections of $e^+e^-\to\gamma X(3872)$ to $e^+e^-\to\omega X(3872)$ is calculated using
\begin{linenomath*}
\begin{equation}\label{eq:ratio-gw}
	\frac{\sigma_{\gamma X(3872)}}{\sigma_{\omega X(3872)}} = \frac{N_{\gamma X(3872)}}{N_{\omega X(3872)}}
    \frac{[\varepsilon(1+\delta)]_{\omega X(3872)}^\mathrm{Ave.}}{[\varepsilon(1+\delta)]_{\gamma X(3872)}^\mathrm{Ave.}}
    \mathcal{B}_{\omega},
\end{equation}
\end{linenomath*}
where $N_{\gamma X(3872)}$ and $N_{\omega X(3872)}$ are the signal yields of the $e^+e^-\to\gamma X(3872)$ and $e^+e^-\to\omega X(3872)$ processes, respectively. $[\varepsilon(1+\delta)]_{\gamma X(3872)}^\mathrm{Ave.}$ and $[\varepsilon(1+\delta)]_{\omega X(3872)}^\mathrm{Ave.}$ are the average of product of the selection efficiencies and ISR factors at c.m.~energies between 4.66 and 4.95 GeV for the two processes. $\mathcal{B}_{\omega}$ is the product of the branching fractions of $\omega\to\pi^+\pi^-\pi^0$ and $\pi^0\to\gamma\gamma$.
Since no obvious $\gamma X(3872)$ signal is observed, we determine the upper limit on the cross section ratio $\sigma_{\gamma X(3872)}/\sigma_{\omega X(3872)}$. 
An unbinned maximum likelihood fit is performed to the $M(\pi^+\pi^- J/\psi)$ distribution from the $\gamma X(3872)$ events. The signal PDF is a MC-simulated shape convolved with a Gaussian function, whose parameters are determined from the fit to the $\gamma_{\text{ISR}}\psi(3686)\to\gamma_{\text{ISR}}\pi^+\pi^- J/\psi$ events. The signal yield of $N_{\omega X(3872)}=24.2\pm5.2$ is obtained in Section~\ref{sec:R}, as shown in Fig.~\ref{fig:simfit-wx}. Thus, the upper limit of $\sigma_{\gamma X(3872)}/\sigma_{\omega X(3872)}<0.23$ at 90\% C.L. is determined using the same method described in Section~\ref{sec:R}. The systematic uncertainty (cf.~Section~\ref{sec:sys-3}) is taken into account in the upper limit.

\section{The cross section measurement\label{sec:xs}}

The Born cross section of $e^+e^-\to\omega X(3872)$, $\omega\chi_{cJ}$ at measured c.m.~energy $\sqrt{s}$ is calculated with
\begin{linenomath*}
\begin{equation}
    \sigma^\text{B}(\sqrt{s})=\frac{N}{\mathcal{L}_\text{int}(1+\delta)\frac{1}{|1-\Pi|^2}\varepsilon\mathcal{B}},
\end{equation}
\end{linenomath*}
where $N$ is the signal yield, $\mathcal{L}_\text{int}$ is the integrated luminosity, $(1+\delta)$ is the ISR correction factor obtained from {\sc kkmc}~\cite{ref:kkmc1,ref:kkmc2}, $\frac{1}{|1-\Pi|^2}$ is the vacuum polarization factor~\cite{vp}, $\varepsilon$ is the detection efficiency, and $\mathcal{B}$ is the product of the branching fractions of the intermediate processes.
\begin{figure*}[htbp]
    \centering
\includegraphics[width=0.32\linewidth]{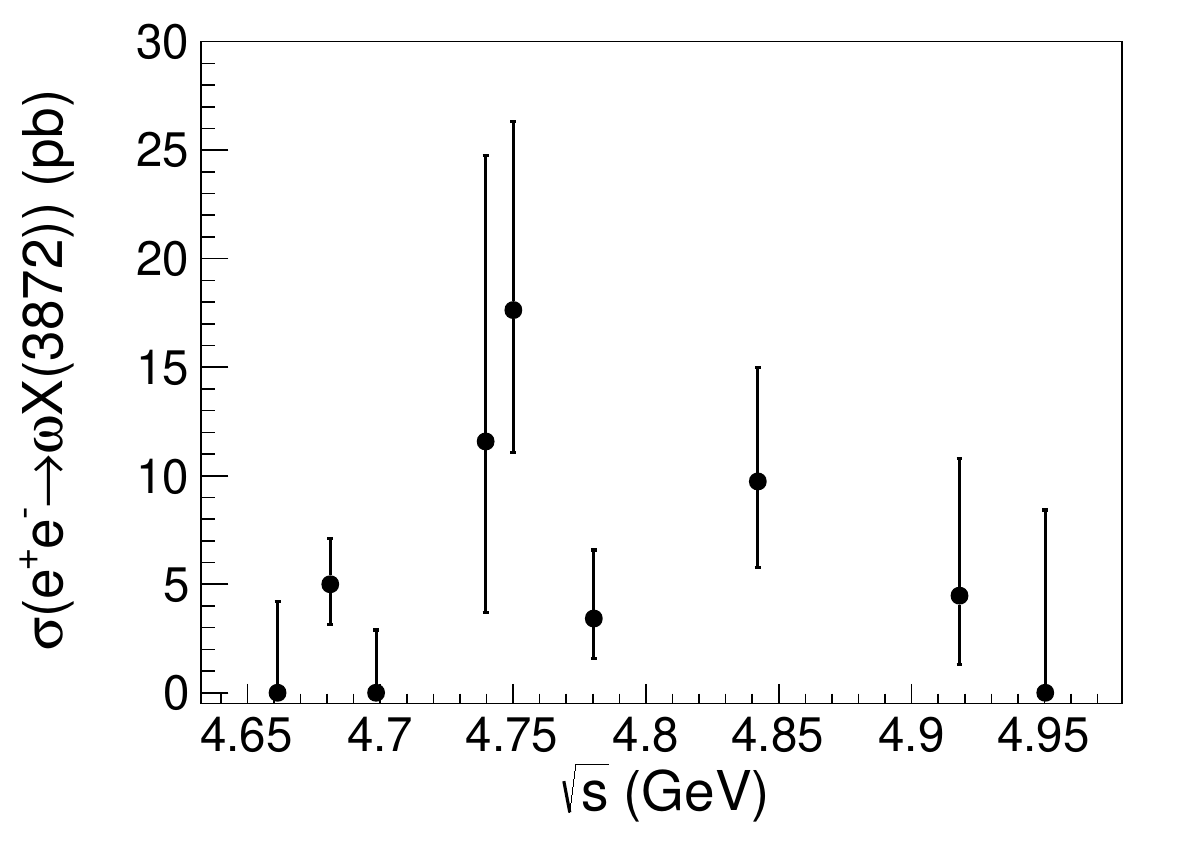}
\includegraphics[width=0.32\linewidth]{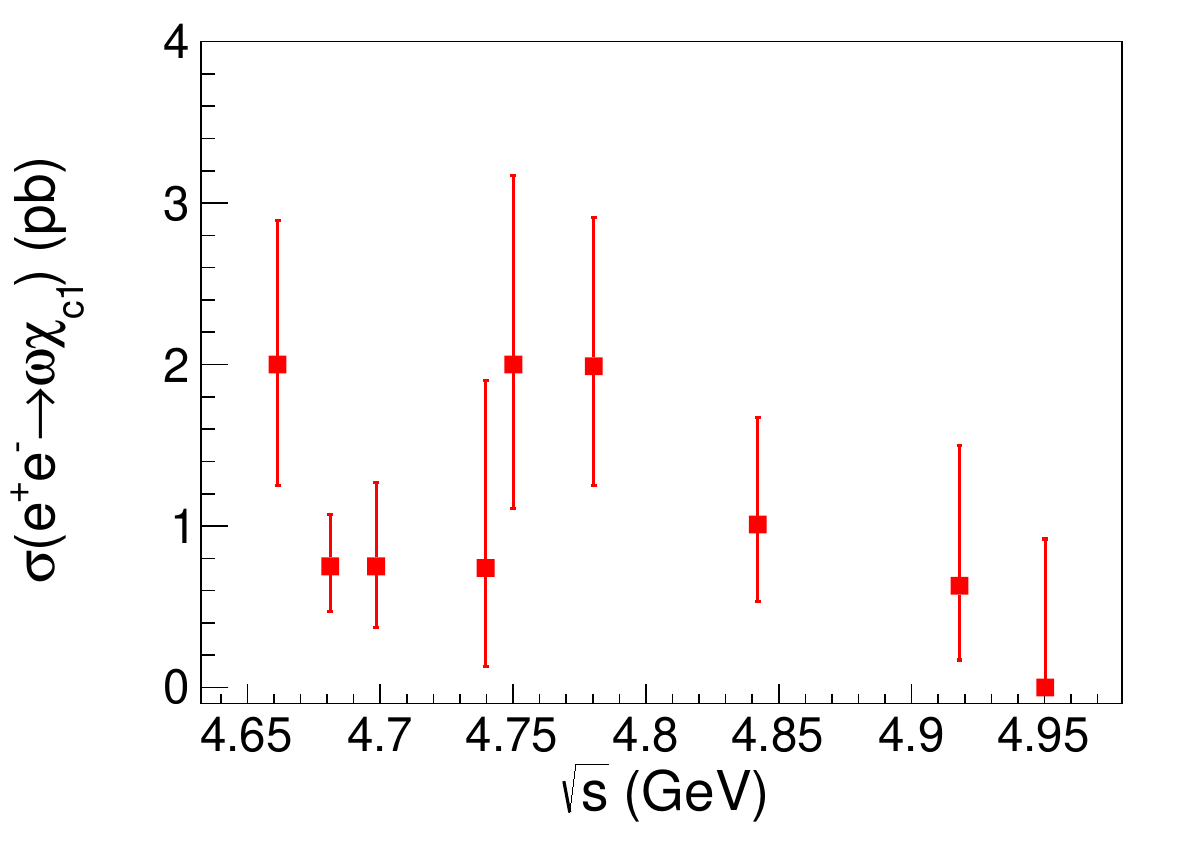}
\includegraphics[width=0.32\linewidth]{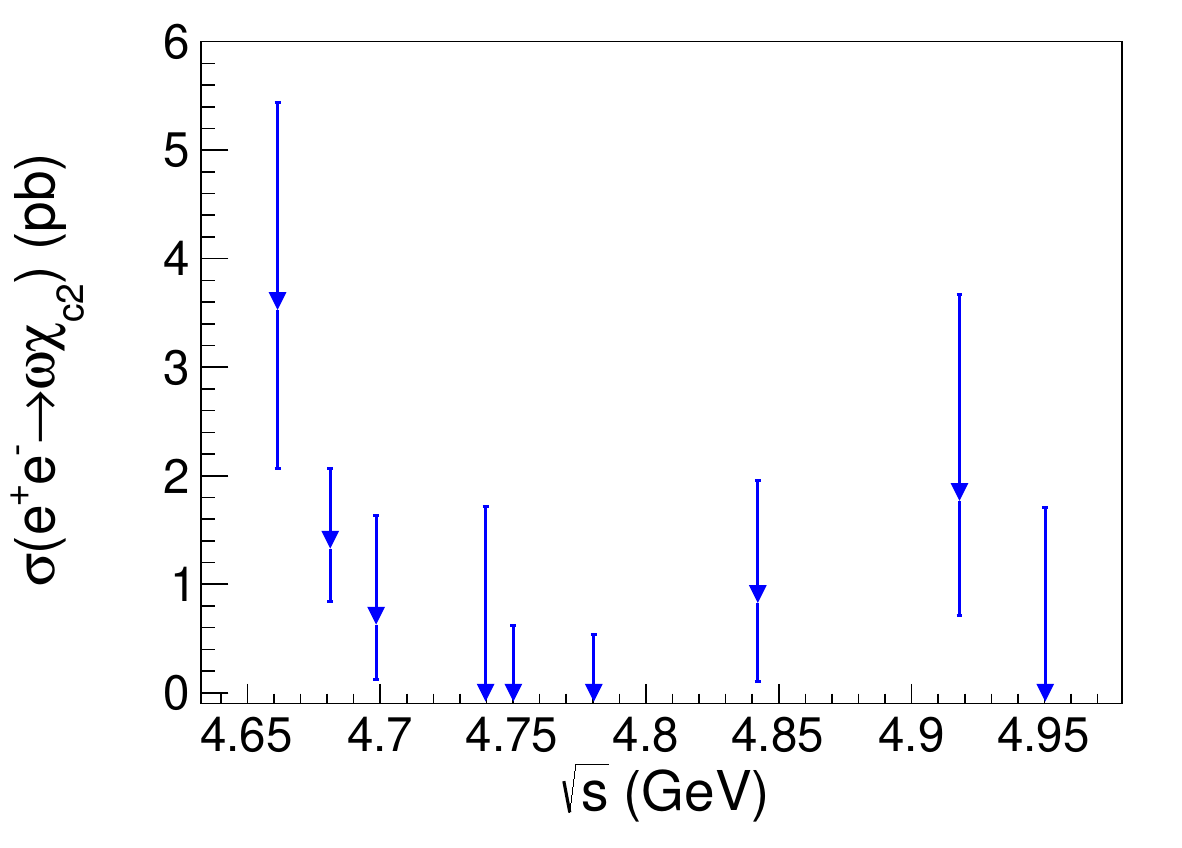}
    \caption{The Born cross sections for the processes of $e^+e^-\to\omega X(3872)$ (left), $e^+e^-\to\omega\chi_{c1}$ (middle), and $e^+e^-\to\omega\chi_{c2}$ (right) at each c.m.~energy. The errors are statistical only.}
    \label{fig:xs}
\end{figure*}

Figure~\ref{fig:xs} shows the cross sections at each c.m.~energy for the processes $e^+e^-\to\omega X(3872)$, $e^+e^-\to\omega\chi_{c1}$ and $e^+e^-\to\omega\chi_{c2}$. The numerical results can be found in Tables~\ref{tab:xsec-wx}, \ref{tab:bxs-c1}, and \ref{tab:bxs-c2}. The fit results at each c.m.~energy are shown in Figs.~\ref{fig:wx} and~\ref{fig:wcj}. 
In the $e^+e^-\to\omega X(3872)$ process, the cross section around 4.75 GeV exhibits a significant increase compared to adjacent energies. This local excess in the cross section near 4.75 GeV is also evident in the $e^+e^-\to\omega\chi_{c1}$ process.

\section{Systematic uncertainty \label{sec:sys}}
\subsection{Systematic uncertainty for $R$ \label{sec:sys-1}}
In the measurement of the ratio $R$, the common systematic uncertainties of $X(3872)\to\gamma J/\psi$ and $X(3872)\to\pi^+\pi^- J/\psi$ cancel out, including the ones due to the luminosity, tracking efficiency of leptons, ISR correction, the muon hit depth in the MUC, and branching fractions of $\omega\to\pi^+\pi^-\pi^0$ and $J/\psi\to\ell^+\ell^-$. The uncommon systematic uncertainties mainly come from the tracking efficiency of pions, photon reconstruction, kinematic fit, MC model, and fit model. 

For the $\omega X(3872)\to\omega\gamma J/\psi$ mode, the final states of  $\pi^+\pi^-\gamma\gamma\ell^+\ell^-$ are reconstructed. 
Taking a systematic uncertainty of 1\% per pion track~\cite{BESIII:2019qvy}, we assign 2\% uncertainty from the pion tracking efficiency. we also assign 2\% uncertainty from photon reconstruction by taking a systematic uncertainty of 1\% per photon~\cite{BESIII:2010ank}. The systematic uncertainty associated with the kinematic fit is estimated by comparing the efficiency difference with and without the correction of the helix parameters of charged tracks in the MC simulations~\cite{BESIII:2012mpj}.  As a result, uncertainties of 1\% and 2.8\% are assigned to the $X(3872)\to\gamma J/\psi$ and $X(3872)\to\pi^+\pi^- J/\psi$ modes, respectively. To estimate the systematic uncertainty from the MC model, the phase space model for the $\omega X(3872)$ and $\gamma  J/\psi$ processes is replaced by a $1\pm \cos^2{\theta}$ distribution. The difference in efficiency is found to be negligible.  The systematic uncertainty related to the fit is investigated by testing various background shapes and changing the fit range. The difference in signal yield is found to be negligible compared to other sources.  

For the $\omega X(3872)\to\omega\pi^+\pi^- J/\psi$ mode, the 5-track ($\pi^+\pi^-\pi^\pm\ell^+\ell^-\gamma\gamma$) and 6-track ($\pi^+\pi^-\pi^+\pi^-\ell^+\ell^-\gamma$) events are reconstructed. According to the previous discussion, we assign a 3\% uncertainty from the pion tracking efficiency and a 2\% uncertainty from the photon reconstruction for the 5-track events. For the 6-track events, the uncertainties from pion tracking efficiency and photon reconstruction are 4\% and 1\%, respectively. The systematic uncertainties from kinematic fit, MC model, and fit are studied using the same method as in the $X(3872)\to\gamma  J/\psi$ mode, with the uncertainties from MC model and fit being negligible.

In the measurement of $R$, the uncertainties from the pion tracking efficiency and the photon reconstruction in the two decay modes of $X(3872)$ partially cancel out. Based on the number of the pion track and photon in the two decay modes, a 1\% uncertainty from the pion tracking efficiency contributes to $R$ for the 5-track case of $\omega \pi^+\pi^- J/\psi$. Similarly, 2\% and 1\% uncertainties respectively from the pion tracking efficiency and photon reconstruction contribute to $R$ for the 6-track case of $\omega \pi^+\pi^- J/\psi$. 
The total uncertainty is obtained by combining the uncertainties of the 5-track ($\Delta_5$) and 6-track ($\Delta_6$) cases $\frac{\varepsilon_5}{\varepsilon_5+\varepsilon_6}\Delta_5+\frac{\varepsilon_6}{\varepsilon_5+\varepsilon_6}\Delta_6$, where $\varepsilon_{5(6)}$ represents the selection efficiency of the 5-track (6-track) events of $\omega\pi^+\pi^- J/\psi$. 

Assuming all the sources of systematic uncertainties are independent, the total systematic uncertainty in $R$ is obtained by adding them in quadrature, as listed in Table~\ref{tab:sys-R}.
\begin{table}[H]
	\renewcommand\arraystretch{1.3}
	\centering
	\caption{The sources of systematic uncertainties and their contributions (in \%) to $R$. \label{tab:sys-R}}
	
	\begin{tabular}{cc}
		\hline\hline
		Source &  Uncertainty \\
		\hline
		Tracking & 1.7 \\

        Photon & 0.7 \\
		
		Kinematic fit in $X(3872)\to\gamma J/\psi$ & 1.0 \\

		Kinematic fit in $X(3872)\to\pi^+\pi^- J/\psi$ & 2.8 \\

		\hline
		Total & 3.5	 \\
		
		\hline\hline
	\end{tabular}
\end{table}

\subsection{Systematic uncertainty for $\sigma_{\omega X(3872)}/\sigma_{\omega \chi_{cJ}}$ \label{sec:sys-2}}

In the measurement of the ratio $\sigma_{\omega X(3872)}/\sigma_{\omega \chi_{cJ}}$, the common systematic uncertainties from luminosity, tracking efficiency of leptons, muon hit depth in the MUC, and branching fractions of $\omega\to\pi^+\pi^-\pi^0$ and $ J/\psi\to\ell^+\ell^-$ in the $\omega X(3872)$ and $\omega \chi_{cJ}$ channels cancel out.

The remaining uncertainties in the $\omega \chi_{cJ}$ process include the tracking efficiency of pions, photon reconstruction, kinematic fit, MC model, ISR correction, branching fraction of $\chi_{cJ}\to\gamma J/\psi$, and the uncertainty of $N_{\omega\chi_{cJ}}$. There are 2\% uncertainty from the tracking efficiency of pions and 2\% uncertainty from photon reconstruction, stemming from the reconstruction of two pion tracks and two photons in the $\omega\chi_{cJ}$ process. The systematic uncertainties from kinematic fit and MC model are estimated using the same method as described in Section~\ref{sec:sys-1}, with the uncertainty of the MC model being negligible. The ISR correction factor and efficiency depend on the input cross section line shape in {\sc kkmc}. By employing the cross section line shapes with and without resonance (here a potential resonance structure around 4.75 GeV in $\omega\chi_{cJ}$ is used) as input, the difference in $[\varepsilon(1+\delta)]^{\text{Ave.}}_{\omega\chi_{cJ}}$ is taken as the systematic uncertainty which is found to be small and can be disregarded. The uncertainty of the branching fraction of $\chi_{cJ}\to\gamma J/\psi$ is taken from the PDG~\cite{pdg}. The statistical uncertainty of $N_{\omega\chi_{cJ}}$ is obtained from the fit as shown in Fig.~\ref{fig:fit-wcj-2d}.

For the $\omega X(3872)$ process, the remaining uncertainties are the tracking efficiency of pions, photon reconstruction, kinematic fit, MC model, ISR correction, branching fraction of $X(3872)\to\pi^+\pi^- J/\psi$ and the uncertainty of $R$.
As both $X(3872)\to\gamma J/\psi$ and $X(3872)\to\pi^+\pi^- J/\psi$ decay modes are considered, these uncertainties are categorized into two groups. The first group contains those with an equal contribution to both modes, such as ISR correction, the branching fraction of $X(3872)\to\pi^+\pi^- J/\psi$, and the uncertainty of $R$. The second group comprises uncertainties that do not equally contribute to the two modes, which include the tracking efficiency of the pion, photon reconstruction, and kinematic fit.

For the first group, the uncertainty of the ISR correction is estimated using the same method as in the $\omega\chi_{cJ}$ process. The uncertainty of the branching fraction of $X(3872)\to\pi^+\pi^- J/\psi$ is quoted from the PDG~\cite{pdg}. In the simultaneous fit of Fig.~\ref{fig:simfit-wx-fixR}, $R=0.25\pm0.06$ is used as an input parameter. By varying the value of $R$ within $\pm 1\sigma$, the difference in the signal yield of $\omega X(3872)$ is taken as the systematic uncertainty. 

The second group of uncertainties cancel out in $\sigma_{\omega X(3872)}/\sigma_{\omega \chi_{cJ}}$ for the $X(3872)\to \gamma  J/\psi$ mode, as the same final states are reconstructed in the processes $\omega X(3872)$ and $\omega \chi_{cJ}$. As for the $X(3872)\to\pi^+\pi^- J/\psi$ mode, these uncertainties in $\sigma_{\omega X(3872)}/\sigma_{\omega \chi_{cJ}}$ are identical to those in $R$ (cf.~Section~\ref{sec:sys-1}), thus we omit the redundant discussions here. To incorporate the second group of uncertainties in the two  decay modes of $X(3872)$, we calculate the weighted average of the uncertainties as
\begin{linenomath*}
\begin{equation}\label{eq:sys-weight}
\Delta^\text{Ave.}=\sum_{i=1}^2\omega_i\Delta_i, \quad \omega_i=\frac{\varepsilon_i\mathcal{B}_i}{\sum_{i=1}^2\varepsilon_i\mathcal{B}_i},
\end{equation}
\end{linenomath*}
where $\Delta^\text{Ave.}$ is the average systematic uncertainty, $\omega_i$, $\Delta_i$, $\varepsilon_i$, and $\mathcal{B}_i$ are the weight, systematic uncertainty, selection efficiency, and branching fraction, respectively, with $i=1$ for the $X(3872)\to\gamma J/\psi$ mode and $i=2$ for the $X(3872)\to\pi^+\pi^- J/\psi$ mode.  

Assuming all the sources of systematic uncertainties are independent, the total systematic uncertainty in the ratio $\sigma_{\omega X(3872)}/\sigma_{\omega \chi_{cJ}}$ is obtained by adding them in quadrature. Table~\ref{tab:sys-xs} summarizes all sources of systematic uncertainties and their contributions to the uncertain of the ratio $\sigma_{\omega X(3872)}/\sigma_{\omega \chi_{cJ}}$.

\begin{table}[H]
	\renewcommand\arraystretch{1.3}

	\centering
	\caption{The sources of systematic uncertainties and their contributions (in \%) to $\sigma_{\omega X(3872)}/\sigma_{\omega\chi_{c1}}$ ($\sigma_{\omega X(3872)}/\sigma_{\omega\chi_{c2}}$). $\Delta_1$ and $\Delta_2$ represent the uncertainties associated with the $X(3872)\to\gamma J/\psi$ and $X(3872)\to\pi^+\pi^- J/\psi$ modes, respectively, their weighted average $\Delta^\text{Ave.}$ is calculated with Eq.~(\ref{eq:sys-weight}).\label{tab:sys-xs}}
  \scalebox{0.9}{

	\begin{tabular}{cccc}
		\hline\hline
		Source &  $\Delta_1$  & $\Delta_2$  & $\Delta^\text{Ave.}$ \\
		\hline
  			
		$\mathcal{B}(\chi_{cJ}\to\gamma J/\psi)$ & \multicolumn{2}{c}{2.9 (2.6)} &  2.9 (2.6)\\
  
        $N_{\omega\chi_{cJ}}$ & \multicolumn{2}{c}{18.6 (30.7)} & 18.6 (30.7)\\

        ISR correction of $\omega X(3872)$ & \multicolumn{2}{c}{3.0} & 3.0 \\
        
		$\mathcal{B}(X(3872)\to\pi^+\pi^- J/\psi)$ & \multicolumn{2}{c}{31.6} & 31.6  \\

		$R$	&	\multicolumn{2}{c}{2.0} & 2.0	\\
		
		Tracking & - & 1.7 & 1.4\\

		Photon & - & 0.7 & 0.6 \\

		Kinematic fit of $\omega \chi_{cJ}$ & - & 1.0 (1.0) & 0.8 (0.8)	\\

		Kinematic fit of $\omega X(3872)$ & - & 2.8 & 2.3 \\

		\hline
		Total & 37.0 (44.3) & 37.1  (44.4) & 37.1  (44.4) \\
		
		\hline\hline
	\end{tabular}
 }
\end{table}

\subsection{Systematic uncertainty for $\sigma_{\gamma X(3872)}/\sigma_{\omega X(3872)}$ \label{sec:sys-3}}
In the measurement of the ratio $\sigma_{\gamma X(3872)}/\sigma_{\omega X(3872)}$, the systematic uncertainties common to both the $\gamma X(3872)$ and $\omega X(3872)$ processes, such as the luminosity, tracking efficiency of leptons, branching fractions of $X(3872)\to\pi^+\pi^- J/\psi$ and $J/\psi\to\ell^+\ell^-$ cancel out. The uncommon systematic uncertainties include the tracking efficiency of pion, photon reconstruction, kinematic fit, ISR correction,  muon hit depth in the MUC, fit and $N_{\omega X(3872)}$.

For the $\gamma X(3872)$ process, the final states of  $\gamma\pi^+\pi^-\ell^+\ell^-$ are reconstructed. Therefore, a 2\% uncertainty from the tracking efficiency of pions and a 1\% uncertainty from photon reconstruction are assigned. The study of the systematic uncertainties from the kinematic fit, ISR correction, and the fit also follows Section~\ref{sec:sys-1}. It is found that the uncertainties from the ISR correction and fit are negligible.  

For the $\omega X(3872)$ process, the uncertainties from the tracking efficiency of pions, photon reconstruction, kinematic fit, and ISR correction have already been discussed in Sections~\ref{sec:sys-1} and~\ref{sec:sys-2}. The systematic uncertainty from the requirement of the muon hit depth in the MUC is studied with the control sample of $e^+e^-\to \mu^+\mu^-$, the difference in efficiency between the data and MC simulation due to the requirement is taken as the systematic uncertainty. The statistical uncertainty of $N_{\omega X(3872)}$ is obtained from the fit as shown in Fig.~\ref{fig:simfit-wx}.

The uncertainties from the tracking efficiency of pions and photon reconstruction in the $\gamma X(3872)$ and $\omega X(3872)$ processes partially cancel out.  Similar to the discussion in Section~\ref{sec:sys-1}, we assign 1\% uncertainty from the tracking efficiency of pions and 1\% uncertainty from the photon reconstruction for the 5-track case of $\omega\pi^+\pi^- J/\psi$. A 2\% uncertainty from the tracking efficiency of pions is assigned for the 6-track case of $\omega\pi^+\pi^- J/\psi$. The uncertainties of the 5-track and 6-track cases are then combined by $\frac{\varepsilon_5}{\varepsilon_5+\varepsilon_6}\Delta_5+\frac{\varepsilon_6}{\varepsilon_5+\varepsilon_6}\Delta_6$.

Assuming all the sources of systematic uncertainties are independent, the total systematic uncertainty in the ratio $\sigma_{\gamma X(3872)}/\sigma_{\omega X(3872)}$ is obtained by adding them in quadrature, as shown in Table~\ref{tab:sys-3}.

\begin{table}[H]
	\renewcommand\arraystretch{1.3}
	\centering
	\caption{The sources of systematic uncertainties and their contributions (in \%) to  $\sigma_{\gamma X(3872)}/\sigma_{\omega X(3872)}$. \label{tab:sys-3}}
	
	\begin{tabular}{cc}
		\hline\hline

		Source &  Uncertainty\\
		\hline
		Tracking & 1.7 \\
		
		Photon & 0.3 \\
		
		Kinematic fit of $\gamma X(3872)$ & 1.3 \\
		
		Kinematic fit of $\omega X(3872)$ & 2.8 \\
		
		ISR correction of $\omega X(3872)$ & 3.0 \\
		
		MUC & 1.9 \\
		
		$N_{\omega X(3872)}$ & 21.5 \\
		
		\hline
		Total & 22.1 \\
		\hline\hline

	\end{tabular}
\end{table}  

\subsection{Systematic uncertainty of the cross section measurement}

The systematic uncertainties in the cross section measurement mainly come from the luminosity, tracking efficiency, photon reconstruction, kinematic fit, ISR correction, muon hit depth in the MUC, branching fractions, and fit. The integrated luminosity is measured using Bhabha events with an uncertainty of 0.6\%~\cite{BESIII:2022ulv}. All other uncertainties are studied in previous sections, and summarized in Table~\ref{tab:sys-wx} for the $e^+e^-\to\omega X(3872)$ process and Table~\ref{tab:sys-wcj} for the $e^+e^-\to\omega \chi_{cJ}$ process.

\begin{table}[H]
	\renewcommand\arraystretch{1.3}
	\centering
	\caption{The sources of systematic uncertainties and their contributions (in \%) to the cross section of $e^+e^-\to\omega X(3872)$. \label{tab:sys-wx}}
	
	\begin{tabular}{cc}
		\hline\hline

		Source &  Uncertainty\\
		\hline
        Luminosity & 0.6 \\
		Tracking & 5.7 \\
		
		Photon & 1.3 \\
				
		Kinematic fit  & 2.8 \\
		
		ISR correction & 3.0 \\
        $\mathcal{B}(J/\psi\to\ell^+\ell^-)$ & 0.6 \\
        MUC & 1.9 \\
		
		\hline
		Total & 7.5 \\
		\hline\hline

	\end{tabular}
\end{table}

\begin{table}[H]
	\renewcommand\arraystretch{1.3}
	\centering
	\caption{The sources of systematic uncertainties and their contributions (in \%) to the cross section of $e^+e^-\to\omega \chi_{c1} (\omega\chi_{c2})$. \label{tab:sys-wcj}}
	
	\begin{tabular}{cc}
		\hline\hline

		Source &  Uncertainty\\
		\hline
        Luminosity & 0.6 \\
		Tracking & 4.0 \\
		
		Photon & 2.0 \\
				
		Kinematic fit  & 1.0 \\
  
		$\mathcal{B}(\chi_{cJ}\to\gamma J/\psi)$ & 2.9 (2.6) \\
	
        $\mathcal{B}(J/\psi\to\ell^+\ell^-)$ & 0.6 \\
        
        MUC & 1.9 \\
		
		\hline
		Total & 5.8 (5.7) \\
		\hline\hline

	\end{tabular}
\end{table} 

\section{Summary}
In summary, with the data sets taken at c.m.~energies from 4.66 to 4.95 GeV, corresponding to an integrated luminosity of $4.5~\text{fb}^{-1}$, we have studied the processes $e^+e^-\to\omega X(3872)$ and $e^+e^-\to\gamma X(3872)$. Through the channel $e^+e^-\to\omega X(3872)$, we have measured the ratio $R\equiv\frac{\mathcal{B}(X(3872)\to\gamma J/\psi)}{\mathcal{B}(X(3872)\to\pi^+\pi^- J/\psi)}$ as $0.38\pm0.20_\text{stat.}\pm0.01_\text{syst.}$ ($R<0.83$ at 90\% C.L.), 
which agrees with Belle's measurement and the previous BESIII's measurement of the $e^+e^-\to\gamma X(3872)$ process. $R$ is currently limited by statistics. 

In addition, using two decay modes of the $X(3872)$, $X(3872)\to\pi^+\pi^- J/\psi$ and $\gamma J/\psi$, we have measured the ratios of average cross sections between 4.66 and 4.95~GeV, $\sigma_{\omega X(3872)}/\sigma_{\omega\chi_{c1}}=5.2\pm1.0_\text{stat.}\pm1.9_\text{syst.}$ and $\sigma_{\omega X(3872)}/\sigma_{\omega\chi_{c2}}=5.5\pm1.1_\text{stat.}\pm2.4_\text{syst.}$. The relatively large cross section for the $e^+e^-\to\omega X(3872)$ process is mainly attributed to the cross section enhancement around 4.75 GeV, which may indicate a potential structure in the cross section of $e^+e^-\to\omega X(3872)$. We have also searched for the process of $e^+e^-\to\gamma X(3872)$ within the same energy range, but have not observed an obvious signal. We set the upper limit on the cross section ratio as $\sigma_{\gamma X(3872)}/\sigma_{\omega X(3872)}<0.23$ at 90\% C.L.. These measurements provide key inputs for a deeper understanding of the production of the $X(3872)$~\cite{Xu:2014zra,Deng:2016stx,Swanson:2004pp,Braaten:2020iye,Ferretti:2014xqa,Eichten:2005ga,Barnes:2005pb,Li:2009zu}. 
 
\begin{acknowledgments}

The BESIII Collaboration thanks the staff of BEPCII and the IHEP computing center for their strong support. This work is supported in part by National Key R\&D Program of China under Contracts Nos. 2020YFA0406300, 2020YFA0406400, 2023YFA1606000; National Natural Science Foundation of China (NSFC) under Contracts Nos. 11635010, 11735014, 11935015, 11935016, 11935018, 12025502, 12035009, 12035013, 12061131003, 12192260, 12192261, 12192262, 12192263, 12192264, 12192265, 12221005, 12225509, 12235017, 12361141819; the Chinese Academy of Sciences (CAS) Large-Scale Scientific Facility Program; the CAS Center for Excellence in Particle Physics (CCEPP); Joint Large-Scale Scientific Facility Funds of the NSFC and CAS under Contract No. U1832207; 100 Talents Program of CAS; Project ZR2022JQ02 supported by Shandong Provincial Natural Science Foundation; Supported by the China Postdoctoral Science Foundation under Grant No. 2023M742100; The Institute of Nuclear and Particle Physics (INPAC) and Shanghai Key Laboratory for Particle Physics and Cosmology; German Research Foundation DFG under Contracts Nos. 455635585, FOR5327, GRK 2149; Istituto Nazionale di Fisica Nucleare, Italy; Ministry of Development of Turkey under Contract No. DPT2006K-120470; National Research Foundation of Korea under Contract No. NRF-2022R1A2C1092335; National Science and Technology fund of Mongolia; National Science Research and Innovation Fund (NSRF) via the Program Management Unit for Human Resources \& Institutional Development, Research and Innovation of Thailand under Contract No. B16F640076; Polish National Science Centre under Contract No. 2019/35/O/ST2/02907; The Swedish Research Council; U. S. Department of Energy under Contract No. DE-FG02-05ER41374



\end{acknowledgments}

\clearpage
\appendix

\begin{table}[H]
	\renewcommand\arraystretch{1.5}
	\centering
	
	\caption{The product of Born cross section of $e^+e^-\to\omega X(3872)$ and the branching fraction of $X(3872)\to\pi^+\pi^- J/\psi$ $\sigma^\text{B}\mathcal{B}$~(pb) at each measured c.m.~energy $\sqrt{s}$~(GeV). The table also includes the integrated luminosity $\mathcal{L}_\text{int}$~($\text{pb}^{-1}$), the detection efficiency $\varepsilon$~(\%), the product of the ISR correction factor and vacuum polarization factor $\frac{1+\delta}{|1-\Pi|^2}$ and the number of the signal yield $N$ at each c.m.~energy. The first uncertainty is statistical and the second systematic. \label{tab:xsec-wx}}
	\begin{tabular}{cccccc}
		\hline
		\hline

	$\sqrt{s}$ & $\mathcal{L}_\mathrm{int}$ & $\varepsilon$ & $\frac{1+\delta}{|1-\Pi|^{2}}$ & $N$ & $\sigma^\mathrm{B}\mathcal{B}$ \\
	
	\hline

	4.66  & 529.63 & 29.01 &  0.50 & $0.0^{+1.3}_{-0.0}$  & $0.0^{+0.16}_{-0.0}\pm0.01$   \\
	
	4.68  & 1669.31 & 28.71 &  0.72 & $6.9^{+3.1}_{-2.4}$  & $0.19^{+0.08}_{-0.07}\pm0.01$   \\
	
	4.70  & 536.45 & 28.07 &  0.74 & $0.0^{+1.3}_{-0.0}$  & $0.0^{+0.11}_{-0.0}\pm0.01$   \\

	4.74  & 164.27 & 29.27 &  0.71 & $1.6^{+1.8}_{-1.1}$  & $0.44^{+0.50}_{-0.30}\pm0.03$   \\
	
	4.75  & 367.21 & 28.91 &  0.81 & $6.1^{+3.0}_{-2.3}$  & $0.67^{+0.33}_{-0.25}\pm0.05$   \\
	
	4.78  & 512.78 & 23.39 &  1.22 & $2.0^{+1.8}_{-1.1}$  & $0.13^{+0.12}_{-0.07}\pm0.01$   \\
	
	4.84  & 527.29 & 14.14 &  1.85 & $5.4^{+2.9}_{-2.2}$  & $0.37^{+0.20}_{-0.15}\pm0.03$   \\
	
	4.92  & 208.11 & 10.55 &  2.56 & $1.0^{+1.4}_{-0.7}$  & $0.17^{+0.24}_{-0.12}\pm0.01$   \\
	
	4.95  & 160.37 & 9.67 &  2.49 & $0.0^{+1.3}_{-0.0}$  & $0.0^{+0.32}_{-0.0}\pm0.02$   \\
	
	\hline
	\hline

	\end{tabular}
\end{table}

\begin{table}[H]
	\renewcommand\arraystretch{1.35}
	\centering
	
	\caption{The Born cross section $\sigma^\text{B}$~(pb) for $e^+e^-\to\omega \chi_{c1}$ at each measured c.m.~energy $\sqrt{s}$~(GeV). The table also includes the integrated luminosity $\mathcal{L}_\text{int}$~($\text{pb}^{-1}$), the detection efficiency $\varepsilon$~(\%), the product of the ISR correction factor and vacuum polarization factor $\frac{1+\delta}{|1-\Pi|^2}$ and the number of the signal yield $N$ at each c.m.~energy. The first uncertainty is statistical and the second systematic. \label{tab:bxs-c1}}
	\begin{tabular}{cccccc}
		\hline
		\hline
		
		$\sqrt{s}$ & $\mathcal{L}_\mathrm{int}$ & $\varepsilon$ & $\frac{1+\delta}{|1-\Pi|^2}$ & $N$ & $\sigma^\mathrm{B}$ \\
		
		\hline

		4.66  & 529.63 & 18.59  &  1.41  & $10.0^{+4.5}_{-3.8}$  & $2.00^{+0.89}_{-0.75}\pm0.12$   \\
		
		4.68  & 1669.31 & 16.04  &  1.69  & $12.3^{+5.2}_{-4.6}$  & $0.75^{+0.32}_{-0.28}\pm0.04$   \\
		
		4.70  & 536.45 & 13.37  &  2.13  & $4.2^{+2.9}_{-2.1}$  & $0.75^{+0.52}_{-0.38}\pm0.04$   \\

		4.74  & 164.27 & 16.62  &  1.25  & $0.9^{+1.4}_{-0.8}$  & $0.74^{+1.16}_{-0.61}\pm0.04$   \\
		
		4.75  & 367.21 & 19.25  &  1.02  & $5.2^{+3.1}_{-2.3}$  & $2.00^{+1.17}_{-0.89}\pm0.12$   \\
		
		4.78  & 512.78 & 20.62  &  1.03  & $7.9^{+3.6}_{-2.9}$  & $1.99^{+0.92}_{-0.74}\pm0.12$   \\
		
		4.84  & 527.29 & 18.27  &  1.31  & $4.6^{+3.0}_{-2.2}$  & $1.01^{+0.66}_{-0.48}\pm0.06$   \\
		
		4.92  & 208.11 & 15.64  &  1.55  & $1.2^{+1.6}_{-0.8}$  & $0.63^{+0.87}_{-0.46}\pm0.04$   \\
		
		4.95  & 160.37 & 14.78  &  1.62  & $0.0^{+1.3}_{-0.0}$  & $0.0^{+0.92}_{-0.0}\pm0.05$   \\
		
		\hline
		\hline

	\end{tabular}
\end{table}

\begin{table}[H]
	\renewcommand\arraystretch{1.35}
	\centering

	\caption{The Born cross section $\sigma^\text{B}$~(pb) for $e^+e^-\to\omega \chi_{c2}$ at each measured c.m.~energy $\sqrt{s}$~(GeV). The table also includes the integrated luminosity $\mathcal{L}_\text{int}$~($\text{pb}^{-1}$), the detection efficiency $\varepsilon$~(\%), the product of the ISR correction factor and vacuum polarization factor $\frac{1+\delta}{|1-\Pi|^2}$ and the number of the signal yield $N$ at each c.m.~energy. The first uncertainty is statistical and the second systematic. \label{tab:bxs-c2}}
	\begin{tabular}{cccccc}
		\hline
		\hline
			
		$\sqrt{s}$ & $\mathcal{L}_\mathrm{int}$ & $\varepsilon$ & $\frac{1+\delta}{|1-\Pi|^2}$ & $N$ & $\sigma^\mathrm{B}$ \\
		
		\hline

		4.66  & 529.63 & 17.54 &  1.32 &  $8.9^{+4.5}_{-3.8}$ &  $3.61^{+1.83}_{-1.54}\pm0.21$  \\
		
		4.68  & 1669.31 & 17.08 & 1.35 &  $11.0^{+5.2}_{-4.4}$ &  $1.41^{+0.66}_{-0.57}\pm0.08$  \\
		
		4.70  & 536.45 & 16.81 &  1.38 &  $1.8^{+2.3}_{-1.5}$ &  $0.71^{+0.92}_{-0.59}\pm0.04$  \\

		4.74  & 164.27 & 15.98 &  1.43 &  $0.0^{+1.3}_{-0.0}$ &  $0.0^{+1.72}_{-0.0}\pm0.10$  \\
		
		4.75  & 367.21 & 16.15 &  1.45 &  $0.0^{+1.1}_{-0.0}$ &  $0.0^{+0.62}_{-0.0}\pm0.04$  \\
		
		4.78  & 512.78 & 15.82 &  1.48 &  $0.0^{+1.3}_{-0.0}$ &  $0.0^{+0.54}_{-0.0}\pm0.03$  \\
		
		4.84  & 527.29 & 15.26 &  1.56 &  $2.3^{+2.7}_{-2.1}$ &  $0.91^{+1.05}_{-0.81}\pm0.05$  \\
		
		4.92  & 208.11 & 14.53 &  1.64 &  $1.8^{+1.8}_{-1.1}$ &  $1.85^{+1.82}_{-1.14}\pm0.11$  \\
		
		4.95  & 160.37 & 13.99 &  1.67 &  $0.0^{+1.3}_{-0.0}$ &  $0.0^{+1.71}_{-0.0}\pm0.10$  \\
		
		\hline
		\hline

	\end{tabular}
\end{table}

\begin{figure*}

\includegraphics[width=0.32\linewidth]{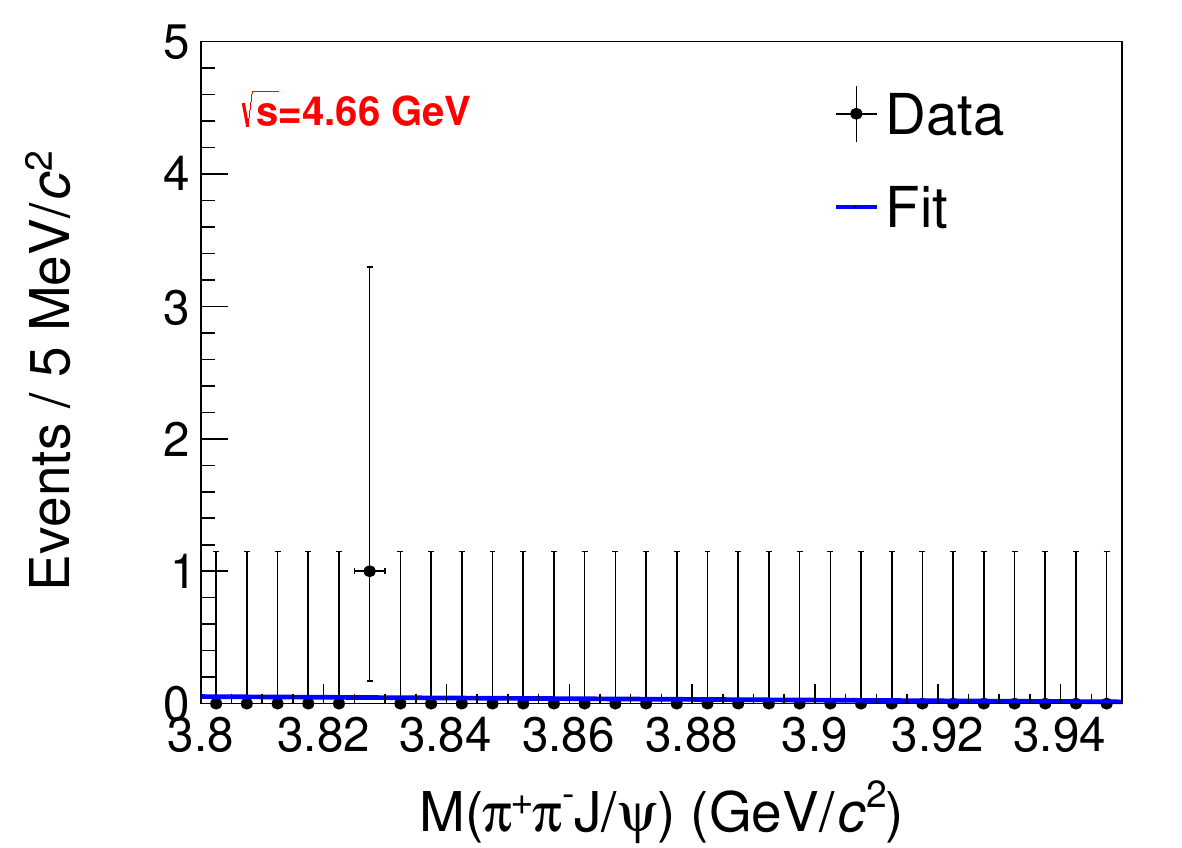}
\includegraphics[width=0.32\linewidth]{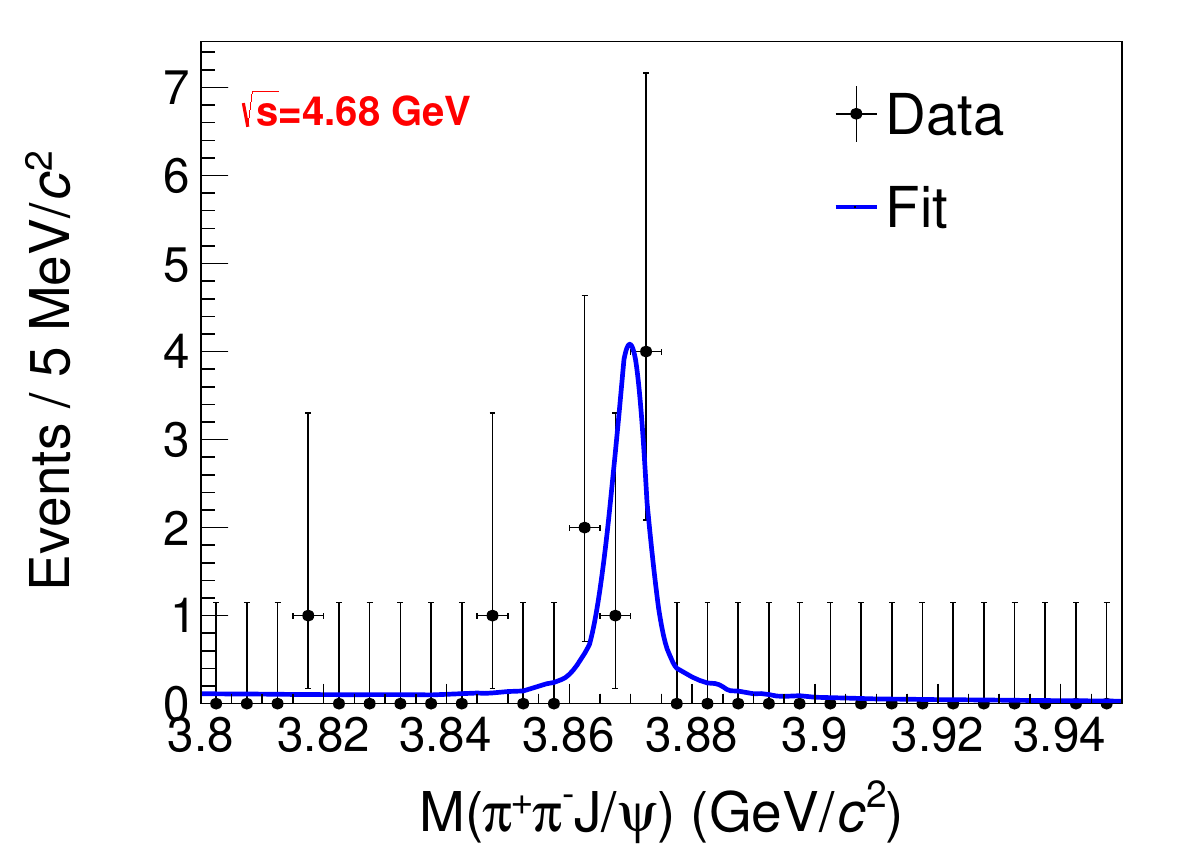}
\includegraphics[width=0.32\linewidth]{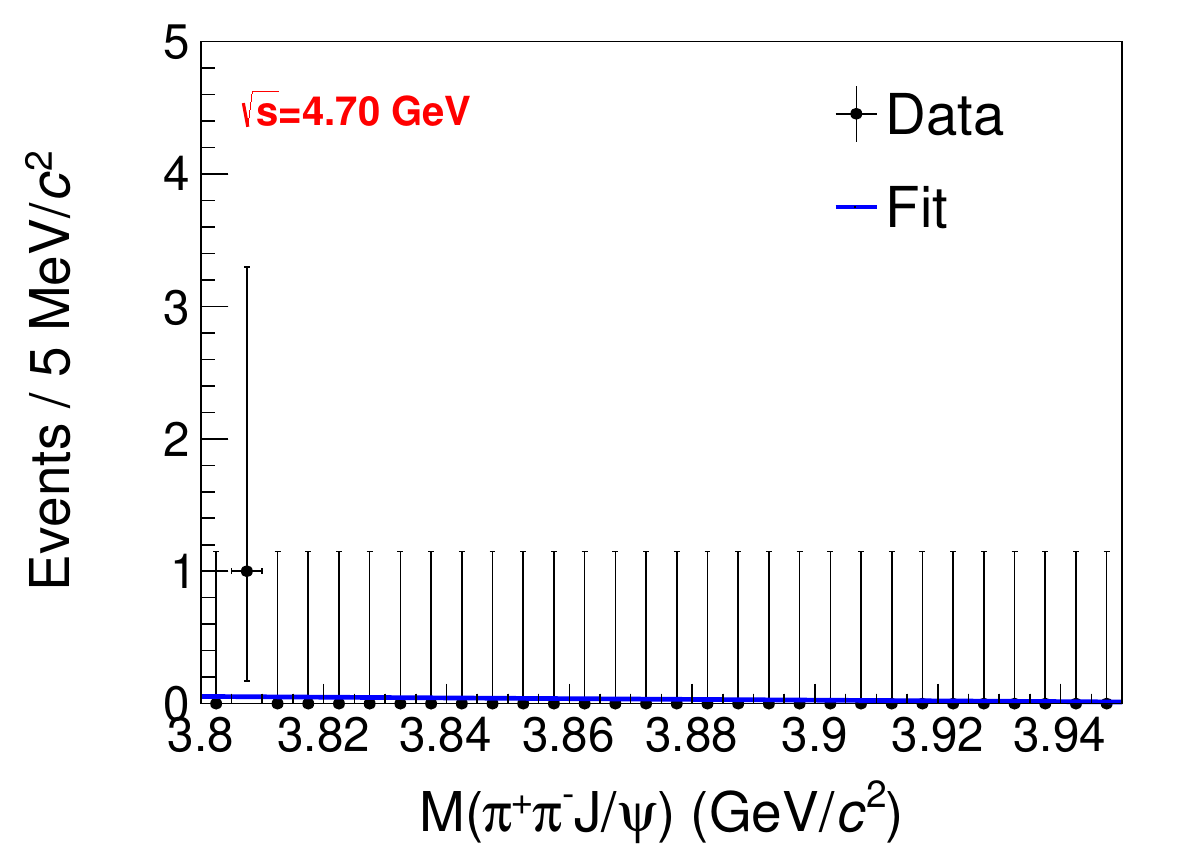}
\vfill
\includegraphics[width=0.32\linewidth]{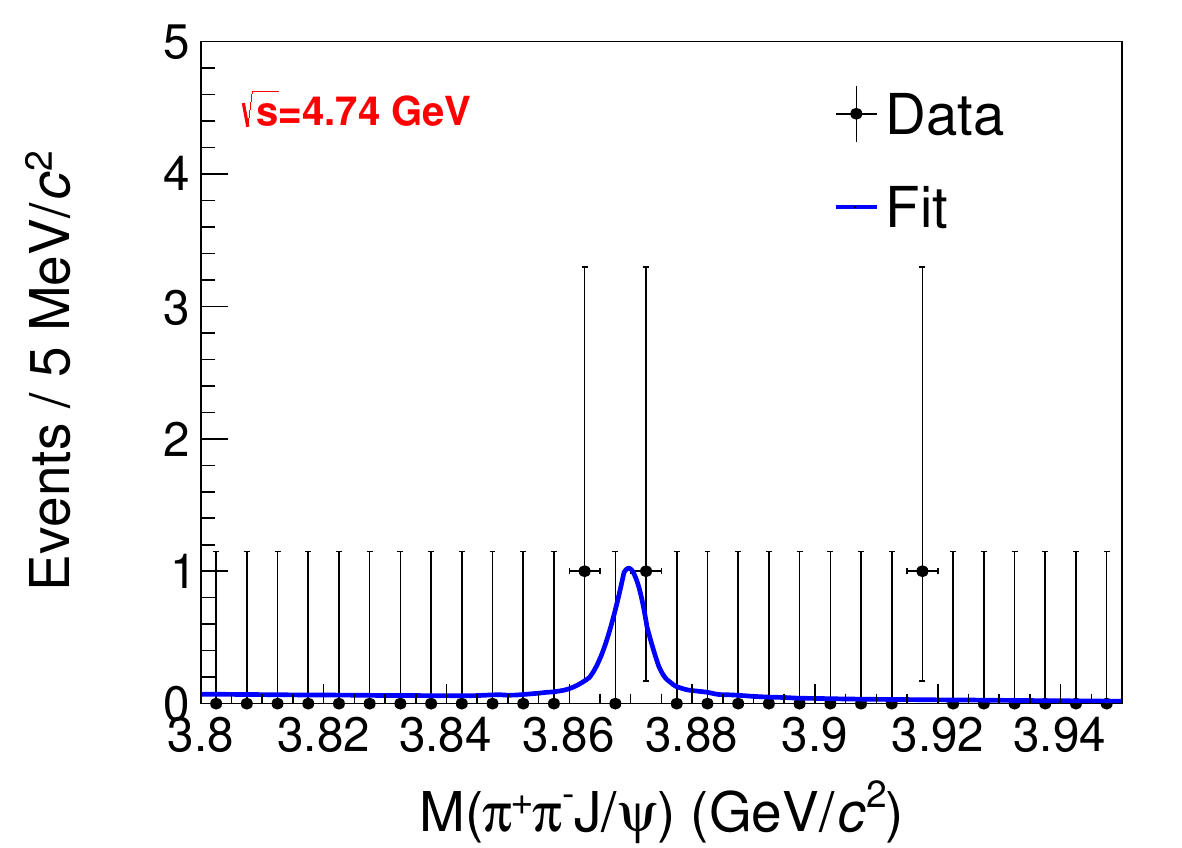}
\includegraphics[width=0.32\linewidth]{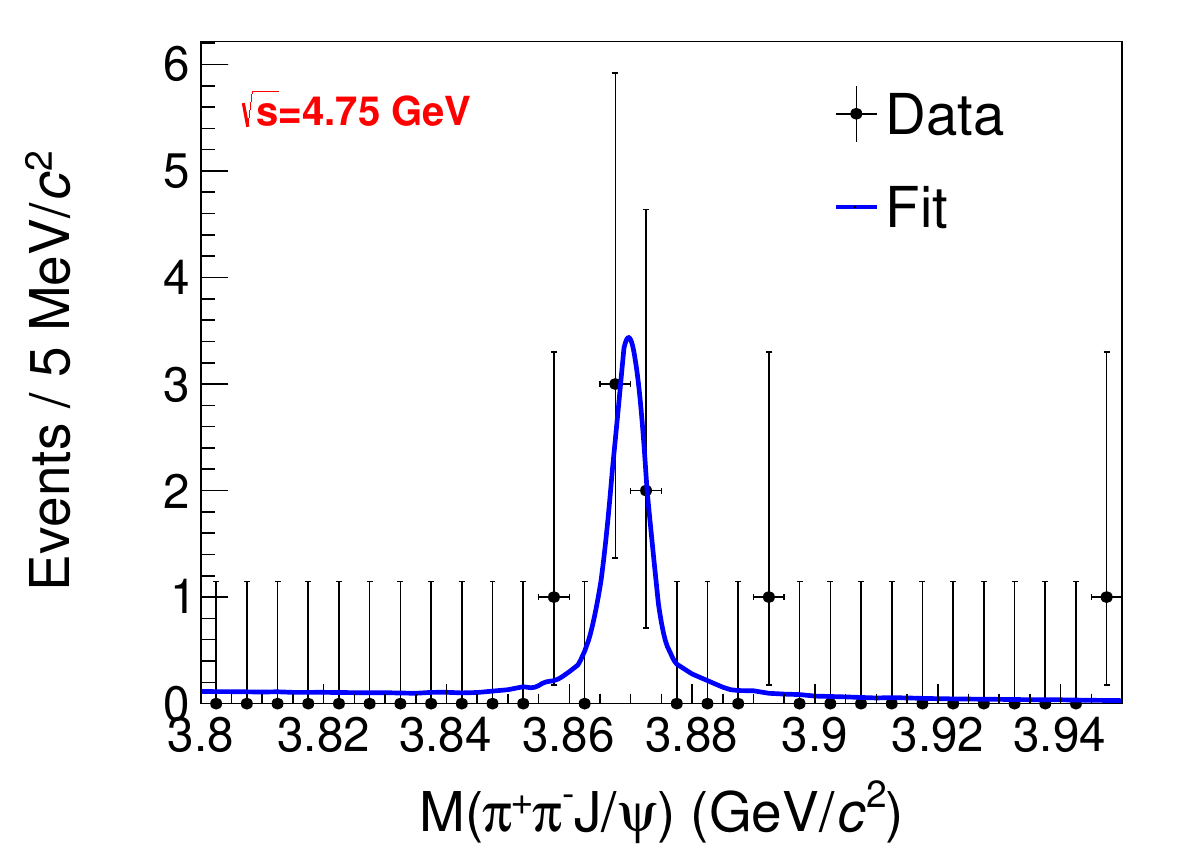}
\includegraphics[width=0.32\linewidth]{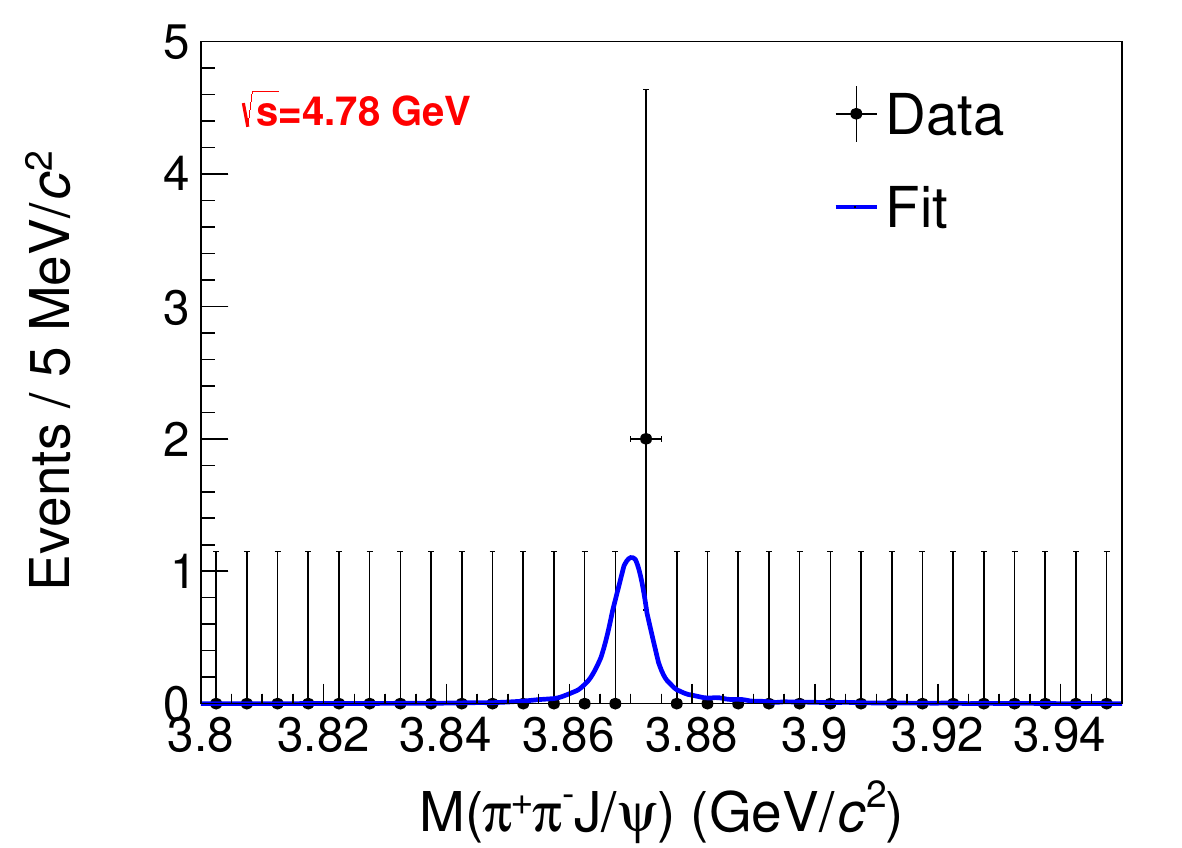}
\vfill
\includegraphics[width=0.32\linewidth]{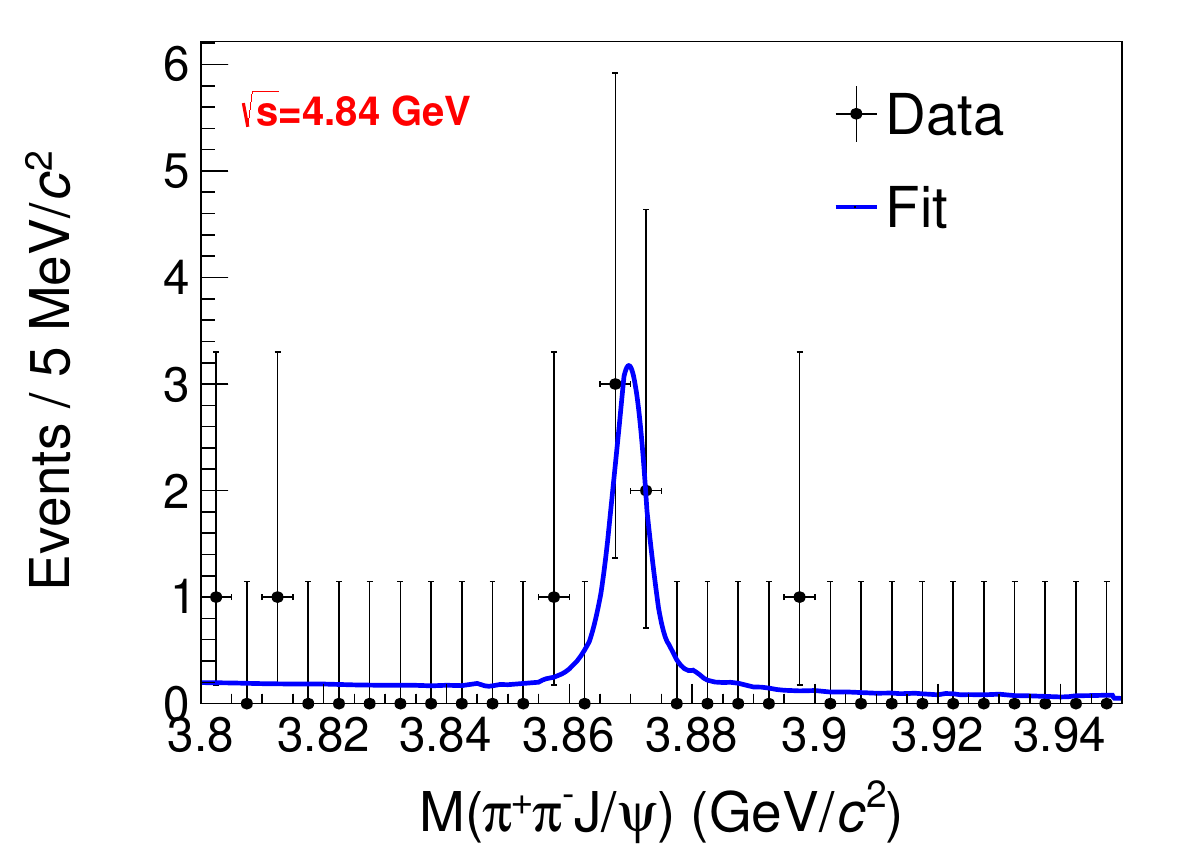}
\includegraphics[width=0.32\linewidth]{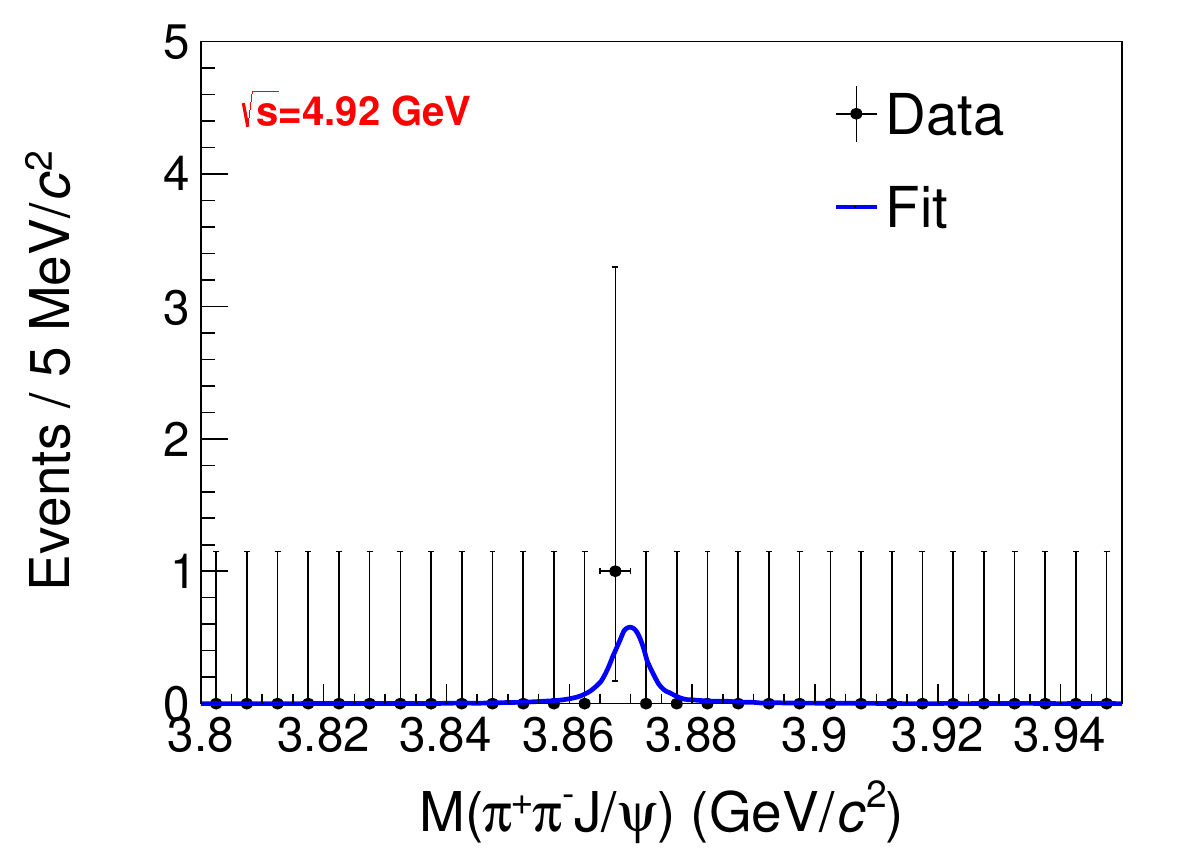}
\includegraphics[width=0.32\linewidth]{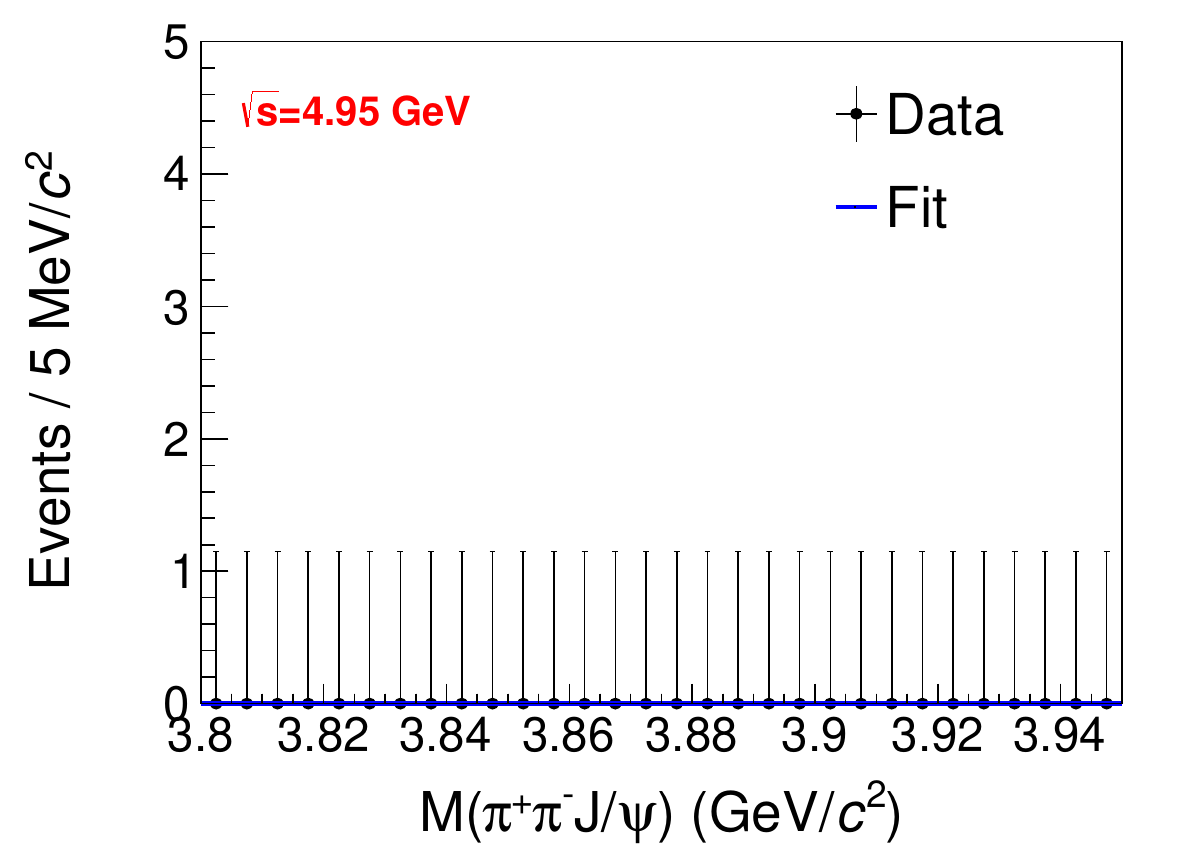}

\caption{The fits to the $M(\pi^+\pi^- J/\psi)$ distributions in the $e^+e^-\to\omega X(3872)$ process at each c.m.~energy. The dots with error bars are data samples, and the solid curves are fit results.}
\label{fig:wx}
    
\end{figure*}

\begin{figure*}
\flushleft
    \includegraphics[width=0.48\linewidth]{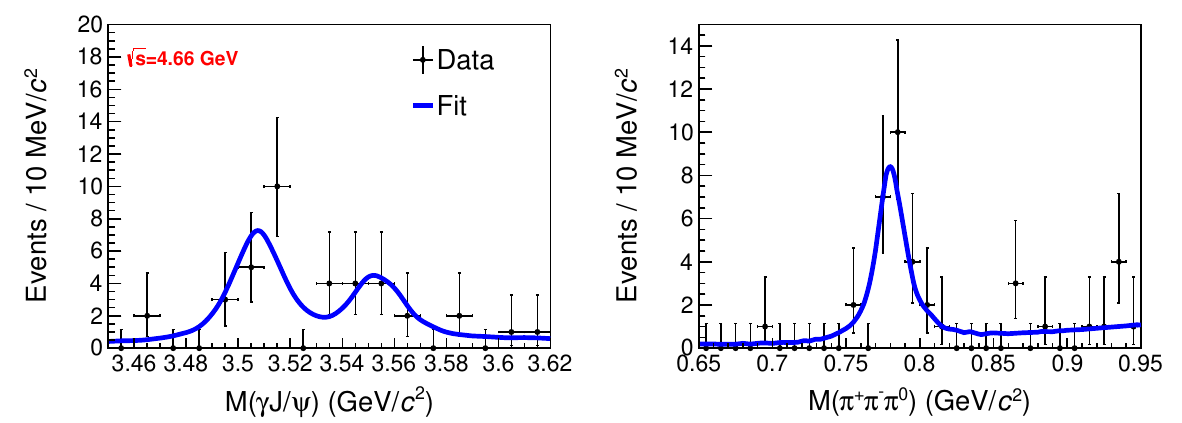}
    \includegraphics[width=0.48\linewidth]{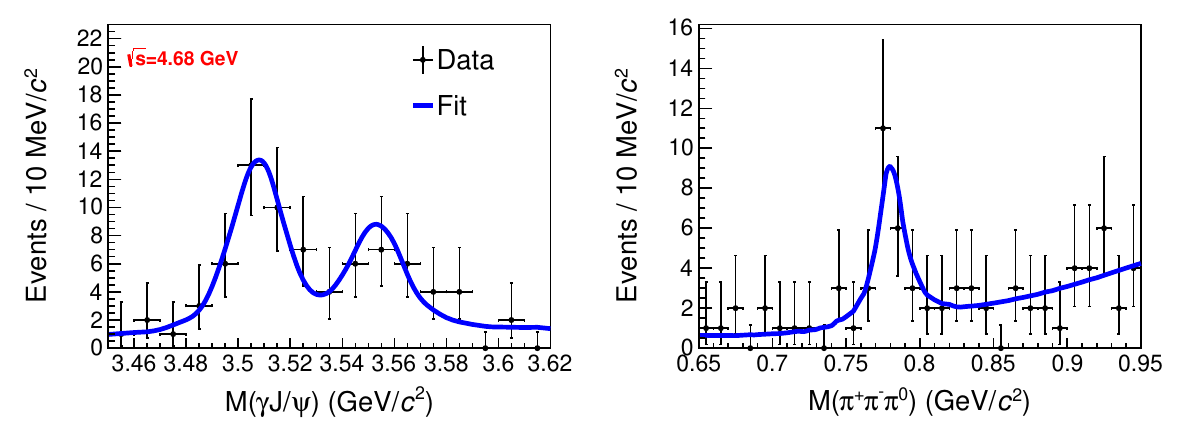}

    \includegraphics[width=0.48\linewidth]{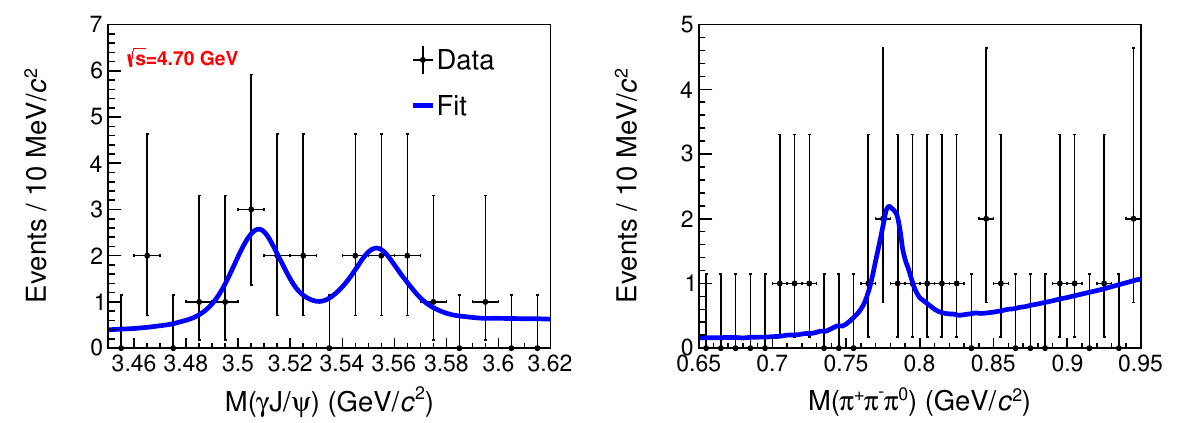}
    \includegraphics[width=0.48\linewidth]{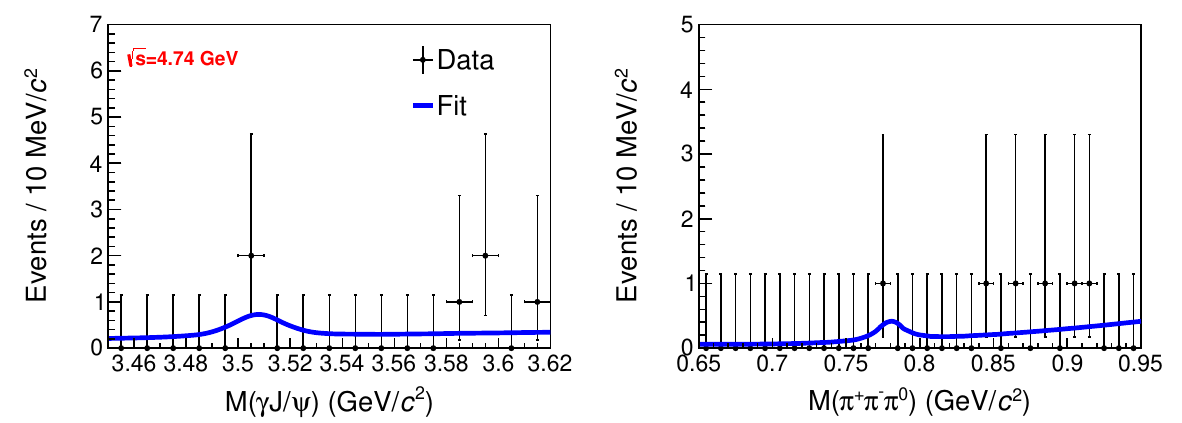}

    \includegraphics[width=0.48\linewidth]{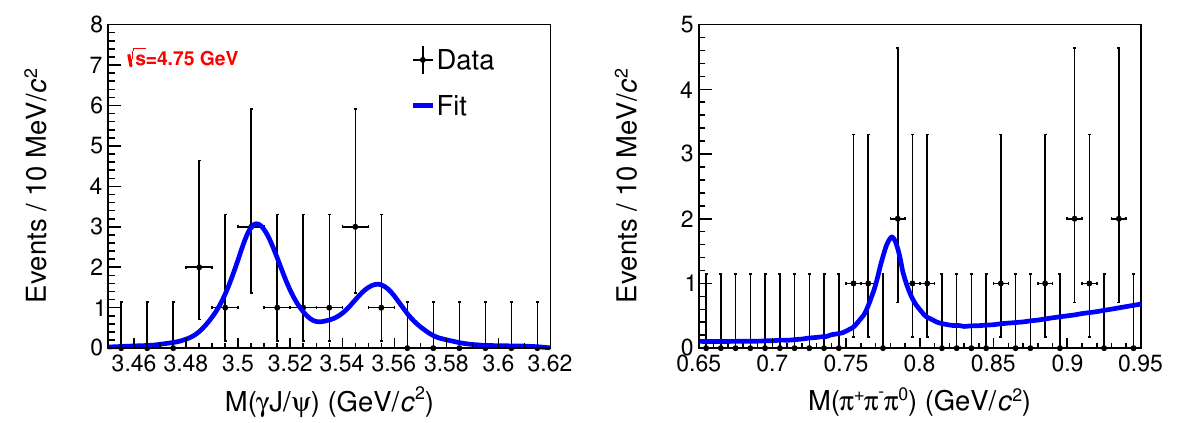}
    \includegraphics[width=0.48\linewidth]{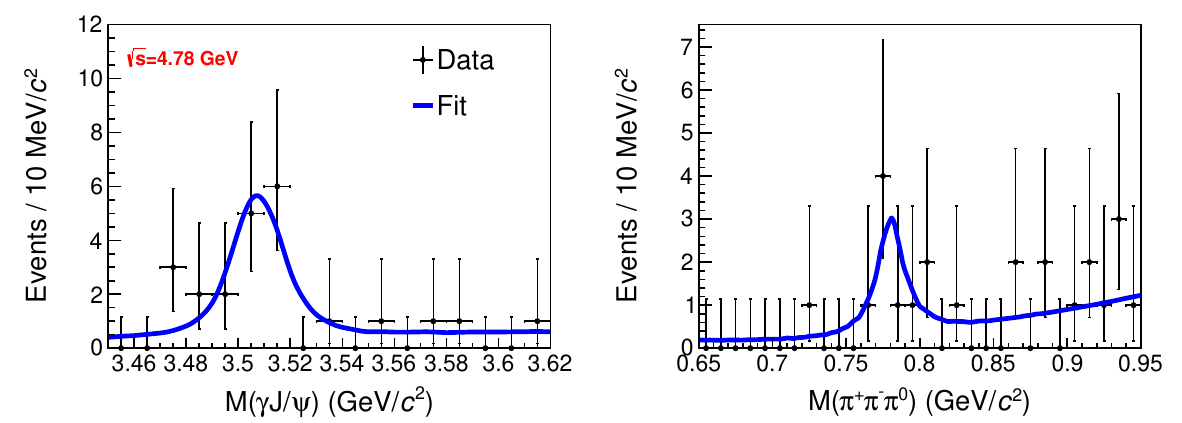}

    \includegraphics[width=0.48\linewidth]{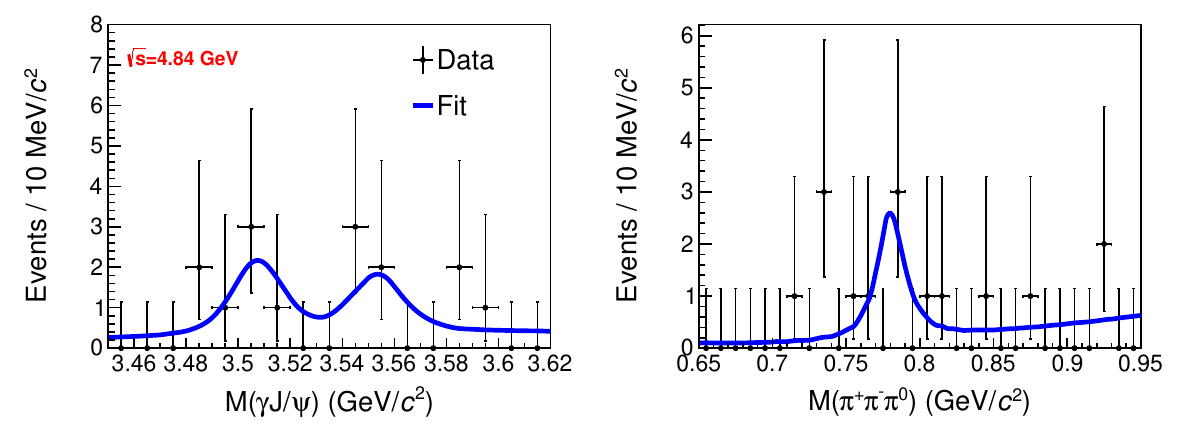}
    \includegraphics[width=0.48\linewidth]{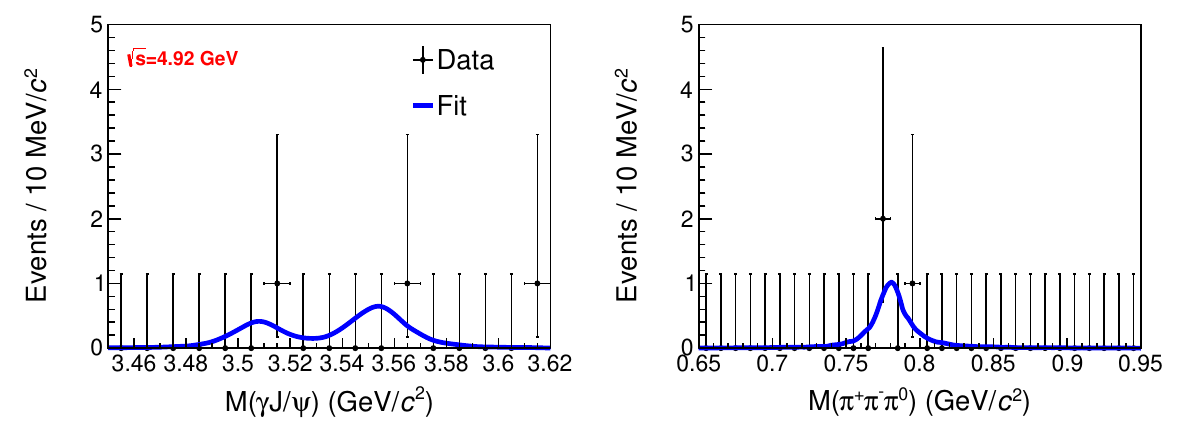}

    \includegraphics[width=0.48\linewidth]{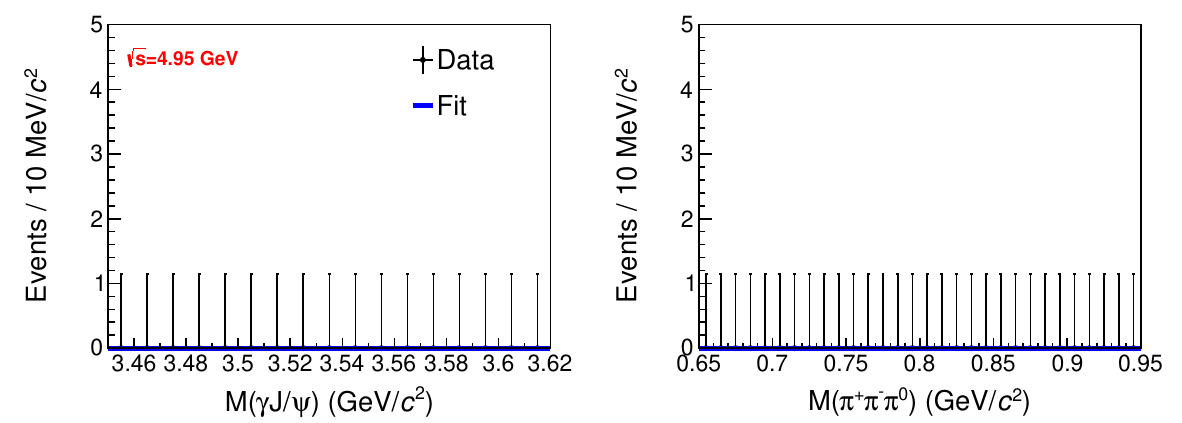}

    \caption{The 2D fits to the $M(\gamma J/\psi)$ and $M(\pi^+\pi^-\pi^0)$ distributions in the $e^+e^-\to\omega \chi_{cJ}$ process at each c.m.~energy. The dots with error bars are data samples, and the solid curves are fit results.}
\label{fig:wcj}

\end{figure*}

\clearpage
\bibliographystyle{apsrev4-2}
\bibliography{bibitem}

\end{document}